\documentclass[twocolumn]{aastex631}

\usepackage{graphicx}
\usepackage{longtable}
\usepackage{ wasysym }

\begin{document}
\title{Photometric White Dwarf Rotation}

\author[0009-0009-6670-0943]{Gabriela Oliveira da Rosa}
\affiliation{Instituto de F\'{\i}sica, Universidade Federal do Rio Grande do Sul \\ 91501-970 Porto Alegre, RS, Brazil\\}
\author[0000-0002-7470-5703]{ S. O. Kepler}
\affiliation{Instituto de F\'{\i}sica, Universidade Federal do Rio Grande do Sul \\ 91501-970 Porto Alegre, RS, Brazil\\}
\author[0009-0006-9702-9484]{L. T. T. Soethe}
\affiliation{Instituto Federal de Educação, Ci\^encia e Tecnologia Sul-rio-grandense \\ 96745-000 Charqueadas, RS, Brazil\\}
\affiliation{Instituto de F\'{\i}sica, Universidade Federal do Rio Grande do Sul \\ 91501-970 Porto Alegre, RS, Brazil\\}
\author[0000-0002-0797-0507]{Alejandra D. Romero}
\affiliation{Instituto de F\'{\i}sica, Universidade Federal do Rio Grande do Sul \\ 91501-970 Porto Alegre, RS, Brazil\\}
\author[0000-0002-0656-032X]{Keaton J.\ Bell}
\affiliation{Department of Physics, Queens College, City University of New York, \\ Flushing, NY, 11367, USA\\}

\begin{abstract}
    
We present a census of photometrically detected rotation periods for white dwarf stars. We analyzed the light curves of 9285 white dwarf stars observed by the Transiting Exoplanet Survey Satellite (\textrm{TESS}) up to sector 69. Using Fourier transform analyses and the \textsc{TESS\_localize} software, we detected variability periods for 318 white dwarf stars. The 115 high probability \textit{likely single} white dwarfs in our sample have a median rotational period of 3.9 hours and a median absolute deviation of 3.5~h.  
Our distribution is significantly different from the distribution of the rotational period from asteroseismology, which exhibits a longer median period of 24.3 hours and a median absolute deviation of 12.0~h.
In addition, we reported non-pulsating periods for three known pulsating white dwarfs with rotational period previously determined by asteroseismology: NGC~1501, TIC~7675859, and G226-29.
We also calculated evolutionary models that include six angular momentum transfer mechanisms from the literature throughout evolution in an attempt to reproduce our findings. Our models indicate that the temperature-period relation of most observational data is best fitted by models with low metallicity, probably indicating problems with the computations of angular-momentum loss during the high-mass-loss phase. Our models also generate internal magnetic fields through the Tayler–Spruit dynamo.
~
\end{abstract}

\section*{Introduction}
\label{intro}

\par The rotation period is an essential parameter for the study of stellar evolution. It plays a significant role in the stellar dynamo mechanism
and is directly related to the mass loss experienced by stars during the asymptotic giant branch \citep[AGB, e.g.,][]{Dominguez1999,Catalan2008,Romero2015,Choi2016}. 
For single stars and stars in binary systems, photometry allows us to detect the rotation period of stars directly and indirectly through asteroseismology \cite[e.g.][]{Corsico19}.

\par Binary stellar systems can exhibit light variability due to different effects. Depending on the geometry of the system and the orientation of the orbital plane as observed from Earth, the light variability can provide the rotational period of stars or the orbital period of the system.
Variability resulting from ellipsoidal variation or reflection in binary systems provides stellar rotation periods, whereas variability from orbital motion indicates the system's orbital period \citep[e.g.][]{Krtika_2023}. However, in the case of spin-orbit coupling, there is no difference between the orbital and rotational periods. The spin-orbit coupling occurs when there is mass loss due to winds, trying to balance the variation in the spin of each star with the angular momentum of the system. 
\cite{1977A&A....57..383Z} demonstrated that the synchronization time between stellar rotation and the orbital period of a binary system is proportional to $P_{\rm orb}^4$, where $P_{\rm orb}$ is the orbital period of the system. This implies that the shorter the orbital period, the faster the synchronization process occurs. \cite{1977A&A....57..383Z}, suggests that systems with orbital periods shorter than 17~days are synchronized. On the other hand, for compact stars such as white dwarfs, \citet{Fuller11,Pablo12,Fuller13,Fuller17,Fuller23} estimate that synchronization may occur only for orbital periods of the order of 1 hour.

\par Single stars, such as the Sun, may also exhibit light variability because of the presence of dark spots and patches on their surface.
These inhomogeneities are typically associated with the presence of a magnetic field or with a variation in the chemical composition across the stellar surface \citep[e.g.][]{Babcock60,2011IAUS..273..249K}.
In this case, the rotation period as well as some harmonics can be present in the frequency spectrum, depending on the number of spots. 

\par White dwarf (WD) stars are the most common end of stellar evolution, constituting the fate of $95-97\%$ of all stars in the Galaxy \cite[e.g.][]{10.1093/mnras/289.4.973, Poelarends_2008, 2010A&A...512A..10S, doi:10.1146/annurev-astro-081811-125534, Woosley_2015, 10.1093/mnras/stu2180}.
As fossils, WDs preserve valuable information on their past evolution, making them essential for studying the structure, evolution, chemical enrichment, and star formation history of our Galaxy \citep{1994A&A...282...86D}. WDs are classified into spectral classes based on the dominant chemical element on their surface. Approximately $80\%$ of all WDs belong to the DA spectral type \cite[e.g.][]{10.1093/mnras/stab2411}, that is, they exhibit H-dominated spectra. The remaining $20\%$ consist mainly of WD stars of the spectral type DB and DO, featuring helium-dominated atmospheres. 

\par Although WDs have been studied since 1915 \citep{Adams15}, their rotation period distribution is still largely undetermined. Ground-based observations cannot easily measure rotation periods because of atmospheric and instrumental limitations, as WDs are intrinsically faint stars. Furthermore, the time gap due to daylight constrains the period range that can be determined \citep{1990ApJ...361..309N}. Long-time baseline observations from space-based telescopes have largely overcome the latter problem. The Kepler spacecraft was launched in $2009$, obtaining excellent data to search for stellar variability \citep[e.g.][]{Molnar16}. 
However, the failure of its second reaction wheel in $2013$ caused significant noise problems at low frequencies. It presented a serious challenge in identifying rotation periods for faint stars. 
The launch of the Transiting Exoplanet Survey Satellite (\textrm{TESS}) in $2018$ \citep{2014SPIE.9143E....O} allowed for a reliable analysis of low frequencies for WD stars. Even with some gaps, the \textrm{TESS} data enabled the detection of photometric rotation periods directly in the frequency spectrum \citep[e.g.][]{Labadie-Bartz2023}.
Finally, photometry analysis also benefits from the high-precision astrometry obtained by the Gaia mission \citep{2018A&A...616A..10G}, which has considerably improved the determination of the parallax and the positions of each star.

\par Stellar rotation rates can also be estimated through spectroscopic observations, by measuring the rotational broadening of the H$\alpha$ line or the Ca\,II K line, as in \citet{2005A&A...444..565B}.
This method has provided rotational velocity estimates for more than 80 WDs, although, for the majority of the sample, it has allowed only the determination of an upper limit. The results indicate that most WDs exhibit rotational velocities lower than 15~km/s, which corresponds to rotation periods longer than 1\,h for a $0.6\,M_\odot$ WD.

\par Another possibility to estimate rotational periods is through asteroseismology \citep[e.g.,][]{keplerale17}, by measuring the separation between the components of multiplets.
\cite{Hermes17} determined the rotation period for twenty WDs, all pulsating hydrogen atmosphere DAVs, using photometric data from the Kepler spacecraft. The authors combined these results with the early estimates compiled by \cite{Kawaler15}, and found a mean rotation period of $35\pm 28$~h for a sample of 40 WDs, indicating a slight tendency to shorter periods for higher masses. 

\par Our work presents the first analysis of the rotation period distribution of a large sample of WD stars based on photometric data from the \textrm{TESS} telescope. We analyze the 21\,832 light curves of the $9285$ white dwarfs observed by \textrm{TESS} from sectors 1 to 69. 
We also compile all known WDs with rotation periods determined by asteroseismology and search for non-pulsating periods using \textrm{TESS} data. We search for stable variability and use the \textsc{TESS\_localize} software to confirm the signal source, as presented in Section~\ref{data}.
In Section~\ref{pulsating}, we present a compilation of all WDs with the rotation period estimated by asteroseismology to date, which we refer to as the seismological sample.
In Section~\ref{results}, we present the rotation period distribution for 318 WDs, analyze its dependence on the stellar parameters, and compare it with the seismological sample. Additionally, we also report the non-pulsating variabilities we detect for three known pulsating WDs from the seismological sample.
In Section~\ref{tayno}, we discuss our theoretical model computations aimed at reproducing our observational results.
 Finally, we present our concluding remarks in Section~\ref{conclusion}. 

\section{Data Analysis\label{data}}

\par We selected all known WDs \citep[e.g.,][]{2000A&AS..143....9W,MWDD,2019MNRAS.486.2169K,2020MNRAS.497..130T,2020ApJ...898...84K,2021MNRAS.507.4646K,2022MNRAS.509.2674G,2023ApJ...944...56A, 2023MNRAS.518.3055O}, including Gaia candidates from \cite{2021MNRAS.508.3877G}, and matched them with the \textrm{TESS} telescope data for objects up to magnitude $G = 17.5$ --- the observational limit that we determined for the white dwarf analysis using the \textrm{TESS} data. This process yielded a sample of 9285 stars, each of which was observed for 1 to 29 sectors. We analyze the available $21\,832$ light curves for all 9285 white dwarfs observed with \textrm{TESS} from sectors 1 (beginning 25 July 2018) to 69 (ending 20 September 2023). All the \textrm{TESS} data used in this article can be found in MAST \citep{https://doi.org/10.17909/t9-st5g-3177, https://doi.org/10.17909/t9-tcn7-7g94, https://doi.org/10.17909/t9-nmc8-f686, https://doi.org/10.17909/t9-yk4w-zc73}.


\par We use the \textsc{Lightkurve} package \citep{lk18} to download the photometric data observed by \textrm{TESS}. We selected data taken with a cadence of 120~s, and 20~s when available, and processed using the \textsc{SPOC} pipeline \citep{2020RNAAS...4..201C}. The \textrm{TESS} telescope observes each sector for $27$~ days, interrupting the observation every $\sim 13$~ days - or less - to send the data to Earth. Due to these gaps in the light curve, we searched for stable variability only for periods up to 13 days. 

\par We combined the light curves of all sectors in which the star was observed and performed a complete Fourier transform (FT) on it. We computed the false alarm probability (FAP) of 1/1000, that is, the limit beyond which peaks on the Fourier transforms have less than one chance in 1000 of being due to noise. We use this detection limit to define whether the observed signal is significant. The FAP was computed by randomizing (Monte Carlo-like simulation) the fluxes at the observed timings, as described in \cite{1993BaltA...2..515K}. If the FT exhibited any peak with an amplitude above the detection limit, we considered its period as a potential variability signal from the WD. When confronted with the detection of multiple harmonic frequencies in the FT, we analyzed the shape of the dips in brightness and examined the light curve folded in phase to determine the variability period. When a noisy light curve hindered the observation of the variability curve, our approach was to select the harmonic peak with the lowest frequency.

\par Following these criteria, we detected variability periods for the 2728 WDs. To avoid including pulsation periods, we checked whether the temperature of the candidate star with periods shorter than 1500~s matched the temperature range of the DOV, DBV, and DAV/ELMV white dwarf instability strips, and also for linear combination of pulsation frequencies. As we are interested here in rotation periods, we excluded periods that matched any of these pulsation criteria. The excluded stars include 64 known pulsators, 207 new pulsators, and 93 pulsator candidates. We also did not include in this work the 173 variability periods from CVs. 
For the remaining 2191, we performed a simple selection using \textsc{TESS\_localize}. We selected those for which the software indicated that for at least for the first observed sector, the WD was the most likely signal source, eliminating 61\% of the targets. In this way, we compile a target list of 845 WDs with a detected non-pulsator variability frequency, which corresponds to 9\% of the initial list of 9285 WDs.This was a preliminary selection. In Section \ref{localize}, we describe the stricter selection criteria applied to this initial sample of 845 WDs using \textsc{TESS\_localize}.

\par We identified within the WDs that make up our sample those with a detected magnetic field, as well as those in binary systems. For magnetic WDs, we first used the classification from the catalog \citet{2023ApJ...944...56A} and the literature for other objects \citep[e.g.,][]{2016yCat....1.2035M, 2000A&AS..143....9W, 2021MNRAS.507.4646K, 2023MNRAS.518.3055O,2020MNRAS.497..130T, 2020ApJ...898...84K,2022MNRAS.509.2674G}. We used the classification of WDs in binary systems from the literature \citep[e.g.][]{2000A&AS..143....9W, MWDD} and also detected new systems by inspecting the Gaia proper motion and parallax data within a box of $120''$ around the white dwarf and by checking the variability shape of the light curve from \textrm{TESS} data.

\par In addition, we compiled all 63 WDs with rotation periods determined by asteroseismology up to now --- which we refer to as seismological sample --- and matched them with the \textrm{TESS} data. 
45 of these pulsating WDs were in our initial list of WDs observed by TESS. For these stars, we inspect all the light curves, searching for reliable non-pulsating periods. To check the presence of a rotation period in the FT, we identified the pulsation frequencies and their linear combination. Our initial sample of 845 WDs with detected non-pulsating periods includes 14 WDs from the seismological sample.

\subsection{Localizing the signal\label{localize}}

\par The \textrm{TESS} telescope offers long-time light curves and high-frequency resolution. However, because of its large plate scale of $21''\times 21''$ per pixel, it exhibits low spatial resolution. Consequently, signals from nearby sources contaminate many \textrm{TESS} light curves. We overcome this problem using the \textsc{TESS\_localize} software \citep{2023AJ....165..141H} to determine whether the detected period arises from the white dwarf or a nearby star. For each star in our target list, we run \textsc{TESS\_localize} for all sectors in which the star was observed, setting the \textit{Principal Components} parameter to 3 \citep[see][]{2023AJ....165..141H}.

\par There are two crucial \textsc{TESS\_localize} output parameters that need to be checked to ensure the quality of the fit: \textit{Height} and \textit{p-value}. 
When \textsc{TESS\_localize} fits the position of the signals, the strength of the signals is represented as the best-fit \textit{Height} parameter. Similarly to the challenge of identifying stars in an image, the signal needs to stand out significantly from the background noise. \citet{2023AJ....165..141H} recommends that the \textit{Height} parameter exceed five times its uncertainty for the localization to be considered significantly above the noise floor.
Another critical parameter is the \textit{p-value}; this number supports the hypothesis that the location of the fit of the signal is consistent with the position of the source. Because we do not reject this hypothesis, the \textit{p-value} should be reasonably high. We adopt the same threshold as \citet{2023AJ....165..239P}: $'p-value' \geq 0.05$. Applying these criteria, we selected only those sectors for which \textsc{TESS\_localize} achieved a significant localization of the signal of interest that is consistent with the location of a known \textit{Gaia} source brighter than $G=18\,$mag. Therefore, since we excluded noisy sectors, we do not need to be concerned about the variation in noise from sector to sector.

\par The final constraint step involves an examination of the \textit{Relative Likelihood}, also an output parameter of \textsc{TESS\_localize}. Essentially, it ranks the proximity of Gaia sources to the localized position of the signal within the pixels. For the same target, \textsc{TESS\_localize} can fit distinct values of \textit{Relative Likelihood} for each available sector of data. So, for analysis purposes, we denoted the mean \textit{Relative Likelihood} over all selected sectors as \textit{Like}. To be consistent, we only consider that the signal originates from the white dwarf if \textsc{TESS\_localize} indicates the white dwarf as the most probable source for all selected sectors.

\par The two-sample Kolmogorov–Smirnov (K-S) test is a robust statistical method used to assess the likelihood of two distinct samples originating from the same population. To ensure that we are not including false positive data in our sample, we performed the two-sample K-S test comparing samples with different thresholds on the selection parameter \textit{Like}. The K-S tests indicated that we would be discarding valuable data if we embraced a \textit{Like} limit higher than $0.75$. Given that, and the fact that the software estimates \textit{Like}\,$\simeq 0.75$ even for confirmed periods from known variables (see \nameref{AppendixA}), we selected all the stars with \textit{Like}$\geq 0.75$.

\par The \textsc{TESS\_localize} analysis with strict conditions reduced by $62\%$ our initial sample of 845 candidate stars. However, the sample of 845 WD is already a consequence of a preliminary selection using \textsc{TESS\_localize} that resulted in a reduction of 61\%. Therefore, in total, \textsc{TESS\_localize} reduced our sample of 2191 WDs with non-pulsating period detected in 85\%.
It is important to point out that, in this second selection, 79$\%$ of the stars discarded by \textsc{TESS\_localize} were rejected because the software was unable to perform a high-quality fit, presenting low 'p-value' and 'height'. Only 21$\%$ were rejected because the software indicated a low probability of the signal having the WD as the source.

\par Our \textit{Complete sample} consists of 318 WDs, including 76 in binary systems and 30 with a detected magnetic fields, counting 1 WD in a binary system that exhibits a magnetic field. Table~\ref{rottab} in \nameref{AppendixD} lists all stars that compose our samples, informing its variability period, amplitude, \textit{Like}, mean \textit{Height} parameter normalized by uncertainty, temperature, mass, and additional information. Table~\ref{rottab} also reports the fraction of sectors whose \textsc{TESS\_localize} fit achieved our quality criteria as `$Q/S$'. White dwarfs confirmed to be in binary systems or with a detected magnetic field can be identified in Table~\ref{rottab} by the column \textit{Info}. The character `+' indicates binarity (e.g. DA+M and WD+pair), while the characters `H' and `P' indicate magnetic field (e.g. DAH and DAP). WDs in binary systems with the designation `WD+pair' were identified through proper motion and parallax, while the ones with the designation `eclipse' were identified by the shape of the light curve variability. 

\par In addition to the sample of 318 WDs, we found that for only $3$ of the $14$ stars in our seismological sample, the WD is also the source of the non-pulsating periods we detected. These three stars are discussed separately in Section~\ref{results2}. Table~\ref{tab3} presents the details of the non-pulsating periods we identified, including amplitude, and quality parameters from the \textsc{TESS\_localize} fit. 

\subsection{Samples}

\par Stars in binary systems have a completely different evolution from single stars if their companions are close enough to interact, affecting their structure. Consequently, variability periods from close binary systems represent a distinct population from single WD rotational periods. Given that, we divided our sample of 318 WDs between those that probably evolved as single stars and those that might have evolved by binary interaction. 

\par Our \textit{Complete sample} comprises 76 WDs in binary systems: 22 identified through parallax and proper motion, 16 identified by the light curve shape, and 38 identified through spectra analysis from literature. The WDs with pairs we identified through parallax and proper motion are all dozens of AU away from their companions; consequently, they probably evolved single stars. The binary systems in our sample identified through spectral observations comprise a WD and a main-sequence star of spectral type M, L, or K.
About 75$\%$ percent of the stars in this type of binary system are far enough from their pair to evolve as single stars \citep{2004A&A...419.1057W}. However, we can not distinguish the specific evolutionary status of each system; thus, we will also assume that the 38 WDs with pairs identified by spectra analysis had their evolution affected by their pairs. Finally, the 16 WDs that show eclipses on their light curves are likely to be close to their pairs, since the periodicities of the eclipses are shorter than 13~days for all cases. Therefore, we select a sample labeled \textit{WDs with close pair} encompassing the 54 WDs in our sample with pairs identified by spectra and eclipses.

\par It is important to recognize that our sample may include more WDs originally in binary systems than those that we detect. WDs with a mass lower than ~0.45~$M_{\astrosun}$ are formed by binary interaction \citep[e.g.][]{1995MNRAS.275..828M, 2007ApJ...671..761K} since it would take longer than the age of the Universe for them to evolve as single stars. This does not necessarily imply that these stars are currently in binary systems, but they certainly originate from such a population. 
Therefore, we classified the stars in our sample with a mass less than 0.45~M$\odot$ as \textit{Potential WDs with Pair}. We also included stars with an effective temperature higher than 40\,000~K based on the fact that WDs with close pairs might exhibit overestimated temperatures due to the reflection effect caused by a brighter companion  \citep[e.g.][]{Schaffenroth23,Steen24}. Once high-temperature WDs are uncommon, we assumed that WDs with temperatures higher than 40\,000~K, an arbitrary choice, are more likely to have a pair than to be a single high-temperature WD. 
Furthermore, we assumed that the remaining stars --- those neither confirmed to have companions nor meet the criteria for potential pairs --- are \textit{Likely single WDs}. 

\par We performed a K-S test comparing the sample of \textit{Potential WDs with pair} with the sample of \textit{WDs with close pairs}; the test indicated that these samples may originate from the same population.
That suggested that the mass and temperature parameters that we used as a criterion to divide our sample of WDs with no confirmed pairs into \textit{Likely single WDs} and \textit{Potential WDs with pair} were well chosen.

\begin{table}[h]
    \centering
    \begin{tabular}{|l|c|c|c|}\hline
    {\bf Sample}& {\bf Stars} & {\bf Mag.} &  {\bf $\bar{P}\pm$MAD (h)}\\ \hline
    \textit{Complete sample} & 318 & 30 & 6.8 $\pm$ 5.2\\ \hline
    \textit{WDs with close pairs} & 54 & 1 & 8.1 $\pm$ 5.4 \\ \hline
    \textit{Potential WDs with pairs} & 149 & 4 & 7.6 $\pm$ 5.2 \\ \hline
    \textit{Likely single WDs} & 115 & 25 & 3.9 $\pm$ 3.5 \\ \hline
    \end{tabular}
    \caption{Compilation of our samples' information: their star number, number of magnetic WD, their median period ({\bf $\bar{P}$}), and median absolute deviation (MAD).}
    \label{samples}
\end{table}

\par  Our \textit{Complete Sample} consists of 318 WDs, including 54 \textit{WDs with close pairs}, 149 \textit{Potential WDs with pairs}, and 115 \textit{Likely single WDs}. Further details of these samples are available in Table~\ref{samples}, which summarizes information on the quantities of magnetic WD in each sample, their median periods $\bar{P}$, and the period median absolute deviation MAD.

\section{Compilation of rotation periods from asteroseismology}
\label{pulsating}

\par This section compiles results from the literature to compare them with our findings. Consequently, this section does not present the results of our work using TESS data.

\begin{figure}[ht]
    \centering
    \includegraphics[width = 0.5\textwidth]{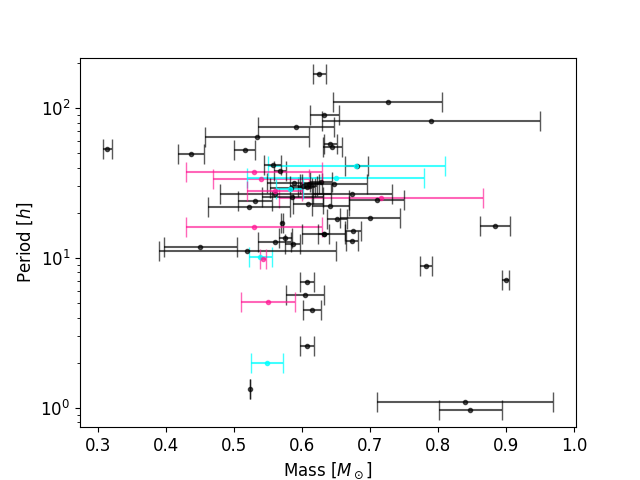}
    \caption{Rotation period versus mass relation, for period determinations from asteroseismology. The period axe is in log scale. Black, cyan, and pink dots indicate DAV, DBV, and DOV stars respectively.}
    \label{massseis}
\end{figure}

\begin{table*}[t!]
    \centering
    \begin{tabular}{|c|c|c|c|c||c|c|c|c|c|} \hline
    Star & $P_{\rm{seism}}$ [h] & Type & Mass [$M_{\astrosun}$] & Ref. &Star & $P_{\rm{seism}}$ [h] & Type & Mass [$M_{\astrosun}$] & Ref.  \\ \hline 


TIC 7675859	& 7.15 &	DAV	& 0.900 $\pm$ 0.005 &   0	&
TIC 55650407 & 13.56 &	DAV	& 0.576 $\pm$ 0.009 &    0	\\ \hline 
TIC 79353860 & 17.06 &	DAV	& 0.571 $\pm$ 0.002 &    0	&
TIC 149863849 & 18.07 &	DAV	& $0.652\pm 0.015$ &    0	\\ \hline 
TIC 230384389 & 22.32 &	DAV	& $0.642\pm 0.027$ &    0	&
TIC 278616553 & 22.81 &	DAV	& $0.609\pm 0.022$ &    0	\\ \hline 
TIC 343296348 & 12.35 &	DAV	& 0.587 $\pm$ 0.011 &    0	&
TIC 353727306 & 4.51 &	DAV	& $0.615\pm 0.013$ &    0	\\ \hline 
TIC 800126377 & 15.1 &	DAV	& $0.675\pm 0.012$ &    0	&
TIC 900762564 & 1.34 &	DAV	& $0.524\pm 0.00$ &    0	\\ \hline 
TIC 1102242692 & 24.14 &    DAV	& $0.531\pm 0.025$ &    0	&
TIC 394015496 	&	29.76	&	DAV	& 0.607 $\pm$ 0.013 &	1 \\ \hline
TIC 21187072 	&	53.76	&	DAV	& 0.3138 $\pm$ 0.0068 &	1	&
PG 1159-035	&	33.6	&	DOV	& 0.54  0.07 &	2	\\ \hline
RX J2117.1+3412 & 25 & DOV & 0.716 0.150 &	3 &
EPIC 228782059 & 34.1 & DBV &  0.65	 0.13 & 4 \\ \hline
GD133	&	168	& DAV & 0.6257	 0.0098 &	5  &
EPIC220274129	&	12.7	&	DAV	& 0.560	 0.024 &	6	\\ \hline
PG 0112+104	&	10.2	&	DBV	& 0.539	 0.017 &	7 &
EPIC220347759	&	31.7	&	DAV	& 0.589	 0.040 & 8  \\ \hline
KIC 4552982	&	18.4	&	DAV	& 0.700	 0.044 &	8 &
KIC 4357037 	&	22	&	DAV	& 0.522	 0.060 &	8  \\ \hline
KIC 7594781 	&	26.8	&	DAV	& 0.674	 0.058 &	8 &
KIC 10132702 	&	11.2	&	DAV	& 0.52	 0.13 &	8  \\ \hline
GD 1212 &	6.9	&	DAV	& 0.608	 0.010 &	8 &
EPIC201719578	&	26.8	&	DAV	& 0.561	 0.082 &	8  \\ \hline
EPIC201730811	&	2.6	&	DAV	&  0.608	 0.010 &	8    &
EPIC201802933	&	31.3	&	DAV	& 0.648	 0.048 &	8  \\ \hline
EPIC210397465	&	49.1	&	DAV	& 0.437	 0.019 &	8    &
EPIC211596649	&	81.8	&	DAV	& 0.79	 0.16 &	8  \\ \hline
EPIC211629697	&	64	&	DAV	& 0.534	 0.076 &	8    &
EPIC211914185	&	1.1	&	DAV	& 0.84	 0.13 &	8  \\ \hline
EPIC211926430	&	25.4	&	DAV	& 0.586	 0.045 &	8    & 
EPIC228682478	&	109.1	&	DAV	& 0.726	 0.080 &	8  \\ \hline
EPIC229227292	&	29.4	&	DAV	& 0.584	 0.031 &	8   &
EPIC220204626	&	24.3	&	DAV	& 0.71 0.04 &	8  \\ \hline
EPIC220258806	&	30	&	DAV	& 0.602	 0.020 &	8    &
EPIC201806008	&	31.3	&	DAV	& 0.611	 0.014 &	8   \\ \hline
KUV02464+3239	&	90.7	&	DAV	& 0.633	 0.021 &	9 &
SDSS J0349-0059	&	9.8	&	DOV	& 0.543 0.004 &	10 \\ \hline
Ross 548	&	37.8	&	DAV	& 0.568	 0.009 & 11 &
GD 165	&	57.3	&	DAV	& 0.642	 0.010 &	11 \\ \hline
WD 1711+657	&	16.4	&	DAV	& 0.884	 0.022 & 12 & 
PG1707+427	&	16	&	DOV	&  0.53 0.1 &	12 	\\ \hline 
G29-38	&	32	&	DAV	& 0.628	 0.016 & 12 &
GD 358	&	29	&	DBV	& 0.582	 0.020 & 12 \\ \hline
EC14012-1446	&	14.4	&	DAV	& 0.632	 0.031 &	12	&
EC20058-5234	&	2	&	DBV	& 0.549	 0.024 &	12	 \\ \hline
KIC 11911480 	&	74.7	&	DAV	& 0.592	 0.056 &	13	&
SDSS J1612+830	&	0.96	&	DAV	& 0.848	 0.047 &	14 \\ \hline
WD 0937+010	&	11.8	&	DAV	& 0.451	 0.053 &	14	&
KUV11370+4222	&	5.7	&	DAV	& 0.605	 0.028 &	15 \\ \hline
HS 0507+0434B	&	40.9	&	DAV	& 0.681	 0.017 &	16 &
GD 154	&	55.2	& DAV & 0.645	 0.014 & 16  \\ \hline
KIC 8626021	&	40.8	&	DBV	& 0.68	 0.13 &	17	&
PG 0122+200	&	37.2	&	DOV	& 0.53 ± 0.1 &	18    \\ \hline   
HL Tau 76	&	52.8	&	DAV	& 0.516	 0.015 &	19	&
G185-32	&	14.5	&	DAV	& 0.6320	 0.0086 &	20	\\ \hline
L19-2	&	13.0	&	DAV	& 0.6742	 0.0088 &	21	&
LP 133-144	&	41.8	&	DAV	& 0.557	 0.012 & 22 \\ \hline 
NGC 1501    &	28.1  &   DOV & 0.56 0.04 &   23  & 
G226-29	&	8.9	&	DAV	& 0.7829	 0.0091 &	24  \\ \hline	
PG 2131+066	&	5.1	&	DOV	& 0.55 0.04 & 	25  &	
	&		&		&  &	\\ \hline	
    \end{tabular}
    \caption{White dwarfs with rotation periods determined previously by asteroseismology.
    References:}  (0)\cite{Romero24}; (1)\cite{Romero22};(2)\cite{2022ApJ...936..187O}; (3)\cite{corsico2021};(4) \citep{2021ApJ...922....2D}; (5) \cite{Fu2019};(6) \cite{Bell_2017}; (7) \cite{Hermes_2017}; (8) \cite{Hermes17}; (9) \cite{Li2017}; (10) \cite{Calcaferro2016}; (11) \cite{Giammichele_2016}; (12) \cite{Kawaler15}; (13) \cite{10.1093/mnras/stt2420}; (14) \cite{Castanheira2013}; (15) \cite{10.1093/mnras/stt2069}; (16) \cite{10.1093/mnras/sts438}; (17) \cite{2011ApJ...736L..39O}; (18) \cite{Fu2007}; (19) \cite{refId0};(20) \cite{Pech2006}; (21) \cite{Bradley_2001}; (22) \cite{teste}; (23) \cite{Bond1996}; (24) \cite{Kepler95}; (25) \cite{1995ApJ...450..350K}. 
    \label{pulsatingtab}
\end{table*}

\par \cite{Kawaler15}, \cite{Hermes17}, and \cite{keplerale17} reported rotation periods of WDs using determinations from asteroseismological analysis. Extending their work, we compile all 63 WD stars with reported seismological rotation periods up to the present. Their rotation periods are listed in Table~\ref{pulsatingtab} as well as their spectroscopy mass and spectral type. Most of the mass reported were taken from \citet{10.1093/mnras/stab2672} because they include the measured parallax.  
The seismological rotational periods have a median rotational period of 24.3~hours and a median absolute deviation of 12.0 h - its histogram is shown in the last line of Figure~\ref{rothist2}. 
Investigating the trend of decreasing rotation period with increasing white dwarf mass claimed by \cite{Hermes17}, we present Figure~\ref{massseis}. The colored dots indicate the spectral types of stars according to the legend. The current plot does not show this trend ; instead, the data appear to be scattered around a central point.

\section{Results}
\label{results}

\par Assuming that any inhomogeneity on the stellar surface will manifest itself as variability with the rotation period of the surface, we attribute the variability periods detected for the \textit{Likely single WDs} to rotation. This consideration follows the exclusion of all periods potentially caused by pulsation. On the other hand, the periods detected for the \textit{WDs with close pairs} and the \textit{Potential WDs with pairs} can represent rotational periods, orbital periods, or even both in case of synchronization. 

\begin{figure}[h!]
    \centering
    \includegraphics[width = 0.47\textwidth,clip]{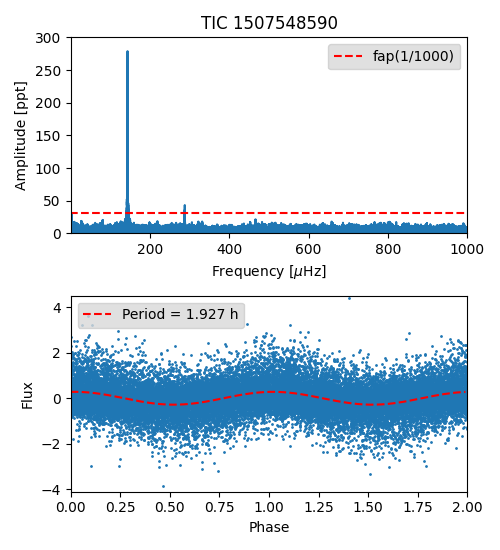}
    \caption{Time series analysis for the star TIC~1507548590. The top panel shows the Fourier Transform as well as its 1/1000 false alarm probability (red dashed line). The bottom panel shows the light curve folded on the period of $1.92772$ hours and a sinusoidal curve with the same period in red.}
    \label{example1}
\end{figure}

\par
\begin{figure}[h]
    \centering
    \includegraphics[width = 0.47\textwidth,clip]{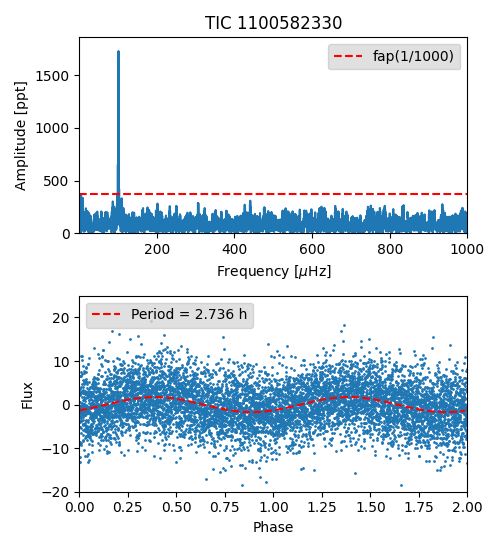}
    \caption{Time series analysis for the star TIC~1100582330. The top panel shows its Fourier Transform as well as its 1/1000 false alarm probability (red dashed line). The bottom panel shows the light curve folded on the period of $2.736$ hours and a sinusoidal curve with the same period in red.}
    \label{example2}
\end{figure}

 Before we present our results, Figures \ref{example1} and \ref{example2} present examples of time-series analysis for two WDs for which we have detected a variability period.
For both figures, the top panel shows the FT with a red dashed line indicating the false alarm probability FAP=1/1000 and the bottom panel shows the folded light curve for the highest period detected in the FT.

Figure~\ref{example1} shows the Gaia WD candidate TIC~1507548590 (SDSS~J$175909.01+231155.7$) that we classified as \textit{Potential WD with pair}.
It shows $T_\mathrm{eff} = 25089\pm 2950$~K and $\log g = 6.404\pm 0.220$~cm\,s$^{-2}$ \citep{2021MNRAS.508.3877G}.
This is a low mass ($0.214~M_\odot$) and faint WD, with magnitude~$G = 17.505$ - our selected faintness limit. The FT presents two significant peaks: a frequency corresponding to $1.92772$~hours and its harmonic. The folded light curve confirms that this is the period corresponding to the main variability of the star. 

\par Figure \ref{example2} shows the DA TIC~1100582330 (SDSS~J$154119.84+120914.6$).
Its spectra fit $T_\mathrm{eff} = 26828\pm 107$~ K and $\log g = 7.60\pm 0.02$~ cm\,s$^{-2}$ \citep{2015MNRAS.446.4078K},
which corresponds to a mass of $0.496~M_\odot$;  it is also faint, at magnitude~$G = 17.032$. The FT presents only one significant peak, with a period of 2.736~hours. This is one of the few stars in our sample of \textit{Likely single WDs} for which it is possible to observe the variability in the folded light curve.

\subsection{Mass and Temperature Distribution \label{massandteff}}

\begin{figure*}[t!]
    \centering
    \includegraphics[width = 1.0\textwidth]{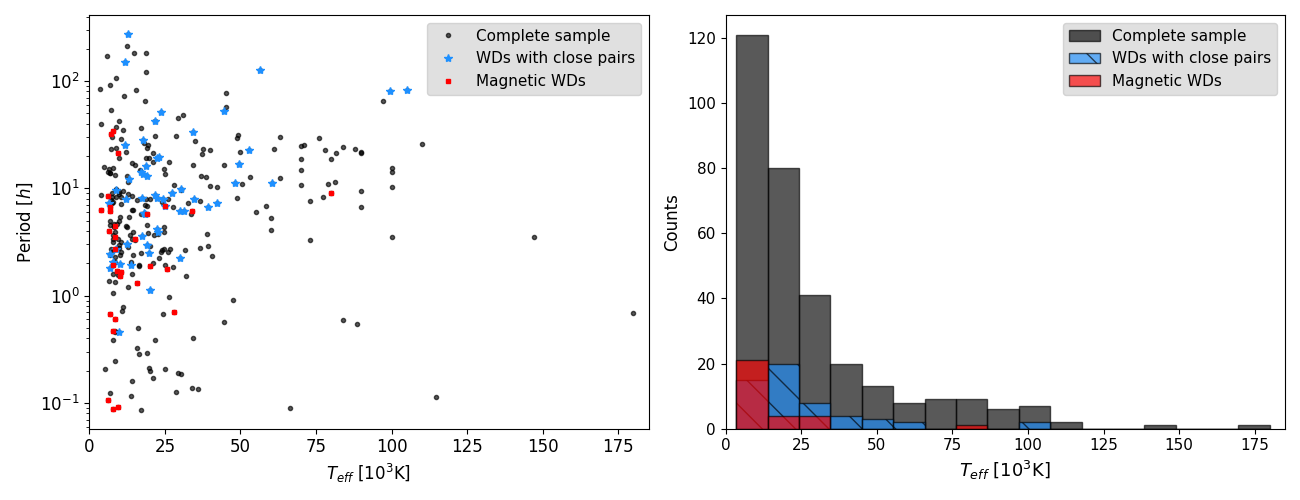}
    \caption{The left panel displays the rotational period in hours versus the effective temperature in Kelvin, both in log scale. The right panel shows the temperature distribution of sample~1. The red and blue dots/bins represent the WDs with magnetic field and with close pairs detected}, respectively.
    \label{teff}
\end{figure*}

\begin{figure*}[t!]
    \centering
    \includegraphics[width = 1.0\textwidth,clip]{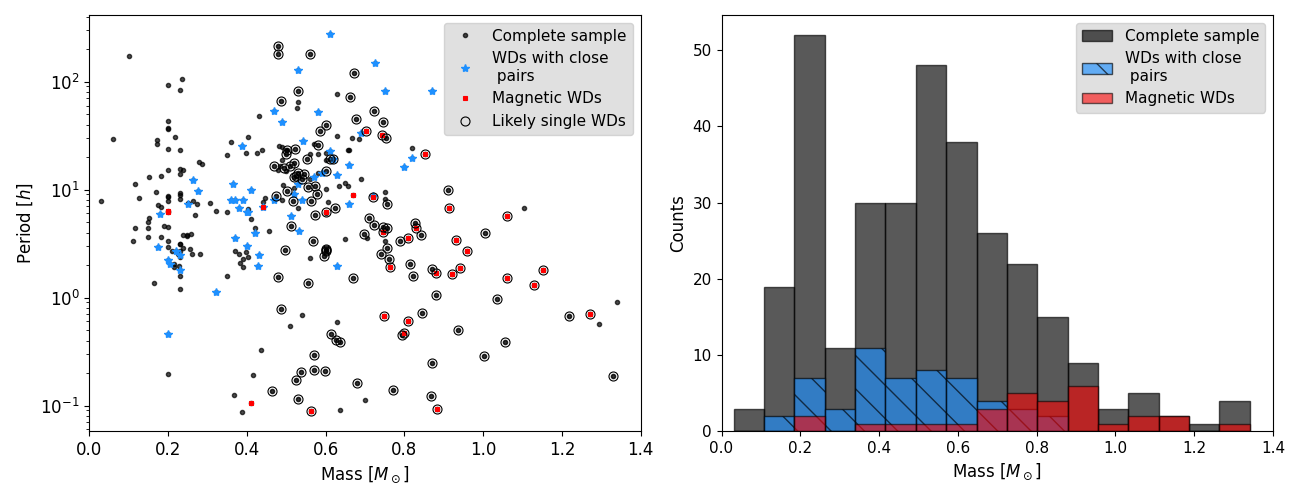}
    \caption{The left panel displays the rotational period in hours in log scale versus mass in solar masses, while the right panel shows the mass distribution of sample~1. The red and blue dots/bins represent the WDs with magnetic field and with close pairs detected}, respectively. The black circles indicate the stars that composes the \textit{Likely single WDs}.
    \label{mass}
\end{figure*}

\par Figures \ref{mass} (\ref{teff}) show the dependence of the detected period --- on a logarithmic scale --- on the stellar mass (effective temperature) in the left panel, and the mass (effective temperature) distribution in the right panel. The red points and bins denote magnetic WDs, whereas the blue points and bins correspond to WDs with close pairs. It is important to point out that the blue and red bins are independent distributions; their overleap does not correspond to a sample of WDs with both close pair and magnetic field detected.

\par Figure \ref{teff} shows that our sample of WDs is distributed around approximately 12,000~K, a representative value of the DA population, which comprises about $80\%$ of the WDs spectroscopically identified \citep[e.g.][]{2021MNRAS.507.4646K}. The left panel of Figure \ref{teff} shows that there does not appear to be a dependence of the detected periods on the effective temperature. Moreover, the close binaries exhibit a mean temperature of 26\,300~K, similar to the \textit{Complete sample}, whose mean temperature is 27\,700~K. On the other hand, magnetic WDs present a trend towards low temperatures, with a mean effective temperature of 17\,100~K.

\par The left panel of Figure~\ref{mass} shows that there is no strong dependence of the detected periods on the stellar mass for the \textit{Complete Sample}. However, this plot reveals a negative correlation for the \textit{Likely single WDs}, suggesting a trend to shorter rotational periods for more massive WDs. Moreover, we can also observe a positive correlation between period and mass for the \textit{WDs with close pairs}.

\par The left panel of Figure~\ref{mass} also shows that most magnetic WDs in our sample exhibit periods shorter than 10~h and that they are among the targets with the shortest periods. It is essential to recognize that our sample has a selection effect. According to \cite{Brinkworth_2013}, the rotational period of magnetic white dwarfs exhibits a bimodal distribution, with one group showing periods on the order of minutes to hours and the other displaying periods much longer than their $4$ years of observations.
However, their total sample consists of only 26 stars, which is smaller than our sample of 30 magnetic WDs.
Each observation cycle (sector) of the \textrm{TESS} telescope lasts $\approx 27$ days, with an interruption of data transfer in the middle, limiting the detection of periods to a few hundred hours. In this way, we detect only the fastest portion of magnetic WDs. Furthermore, massive WDs are fainter and tend to rotate more rapidly \citep[e.g.,][]{Hermes17, Caiazzo21, Williams22}. Although \textrm{TESS} is a small telescope, in fact, most of our magnetic sample consists of massive white dwarfs rotating with periods shorter than $10$~h.

\par The histogram in Figure~\ref{mass} shows that the mass distribution exhibits two peaks, one sharp peak centered on $0.2~M_{\astrosun}$ and a broader distribution around approximately $0.55~M_{\astrosun}$. The distribution around ~0.55~$ M_{\astrosun}$ is consistent with the values expected for the mass distribution of WDs \citep[e.g.][]{2023MNRAS.518.3055O}. On the other hand, the distribution around ~0.2~$\, M_{\astrosun}$, that contains $16\%$ of the sample, cannot be explained by the evolution of single stars.

\par We included in this work all WDs that the \textrm{TESS} telescope was able to observe with sufficient signal-to-noise ratio --- mostly WDs with apparent magnitudes lower than approximately 17.5. Therefore, our sample is limited by magnitude rather than volume. As the mass of WDs is inversely proportional to their radius and the brightness of a star is proportional to its radius, the brighter the WD the less massive it is. This leads to magnitude-limited samples of WDs being biased towards low-mass stars. 

\par
\begin{figure}[h]
    \centering
    \includegraphics[width = 0.47\textwidth,clip]{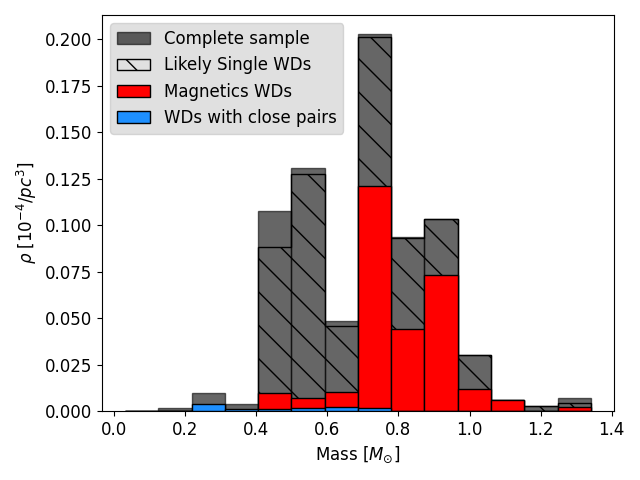}
    \caption{Mass distribution for our \textit{Complete sample} corrected by the $1/V_{max}$ volume. The red and blue bins represent the WDs with magnetic field and with close pairs, respectively. The hatch-filled bins indicated the \textit{Likely single WDs}.}
    \label{masscorrected}
\end{figure}

 To overcome this bias, we corrected the mass distribution of our sample using the method of \citet{Schmidt68} \cite[see e.g.][]{Liebert05,kepler07}. In this method, the contribution of each star is weighted by $1/V_{max}$, where $V_{max}$ is the volume defined by the maximum distance at which the star would still be visible. To correct for the nonuniform distribution of stars in our Galaxy, we assumed a Galactic disk scale height of 250~$pc$, following \cite{1986ApJ...308..176F} and
\cite{Liebert05}. 
We applied this method using parallax, apparent magnitude, and galactic latitude from Gaia DR3 \citep{2018A&A...616A..10G}. Gaia parallax was not available for only one star, which was not included in Figure \ref{masscorrected}.

\par Figure~\ref{masscorrected} exhibits the mass distribution of our sample corrected by the $1/V_{max}$ volume. It shows that the distribution of low-mass WDs around 0.2~$M_{\odot}$ has disappeared, confirming that the excess of brighter low-mass WDs in our sample is an observational bias. Furthermore, the distribution also indicates a lack of WDs with approximately 0.6~$M_{\odot}$, which was supposed to be the most common mass for WD stars. Instead, there is an excess of massive WDs in our sample. Furthermore, it shows that the density of our Complete sample is dominated by the \textit{Likely single WDs} sample
and that the magnetic WDs of our sample dominate the density of the massive WDs. These results could indicate that our \textit{Likely single WDs} sample is dominated by magnetic WDs resulting from binary interactions \cite[e.g.][]{Toonen17,Kilic23}.

\par It is relevant to point out that our initial sample of 845 WDs does not present this lack of 0.6~$M_{\odot}$ stars (see \nameref{AppendixB}). Instead, it exhibits a mass distribution corrected by $1/V_{max}$ as expected for WDs: centered around 0.6.~$M_{\odot}$ \citep[e.g.]{2021MNRAS.507.4646K, 2023MNRAS.518.3055O}. Therefore, this gap only appears after the high-quality  \textsc{TESS\_localize} selection.

\par  As already explained in \nameref{intro}, some inhomogeneity at the stellar surface of single WDs must be present for us to be able to detect their rotational rate through photometry. The most common way of generating inhomogeneities is by magnetic field \cite[e.g.]{2024A&A...683A.227M}. The origin of the WDs magnetic field is still an open question; however, it is well-known that interactions can result in WDs with strong magnetic fields \citep{2008MNRAS.387..897T}. One possibility is that \textsc{TESS\_localize} high-quality parameters may have selected the targets with higher amplitude variability and, consequently, those with higher magnetic fields. One piece of evidence is that \textsc{TESS\_localize} criteria reduced our initial sample by 62$\%$; however, it reduced only by 11.8$\%$ our initial sample of 34 known magnetic WDs.

\subsection{Variability Period Distributions} 
\label{rot}

\begin{figure*}[t!]
    \centering
    \includegraphics[ width = \textwidth]{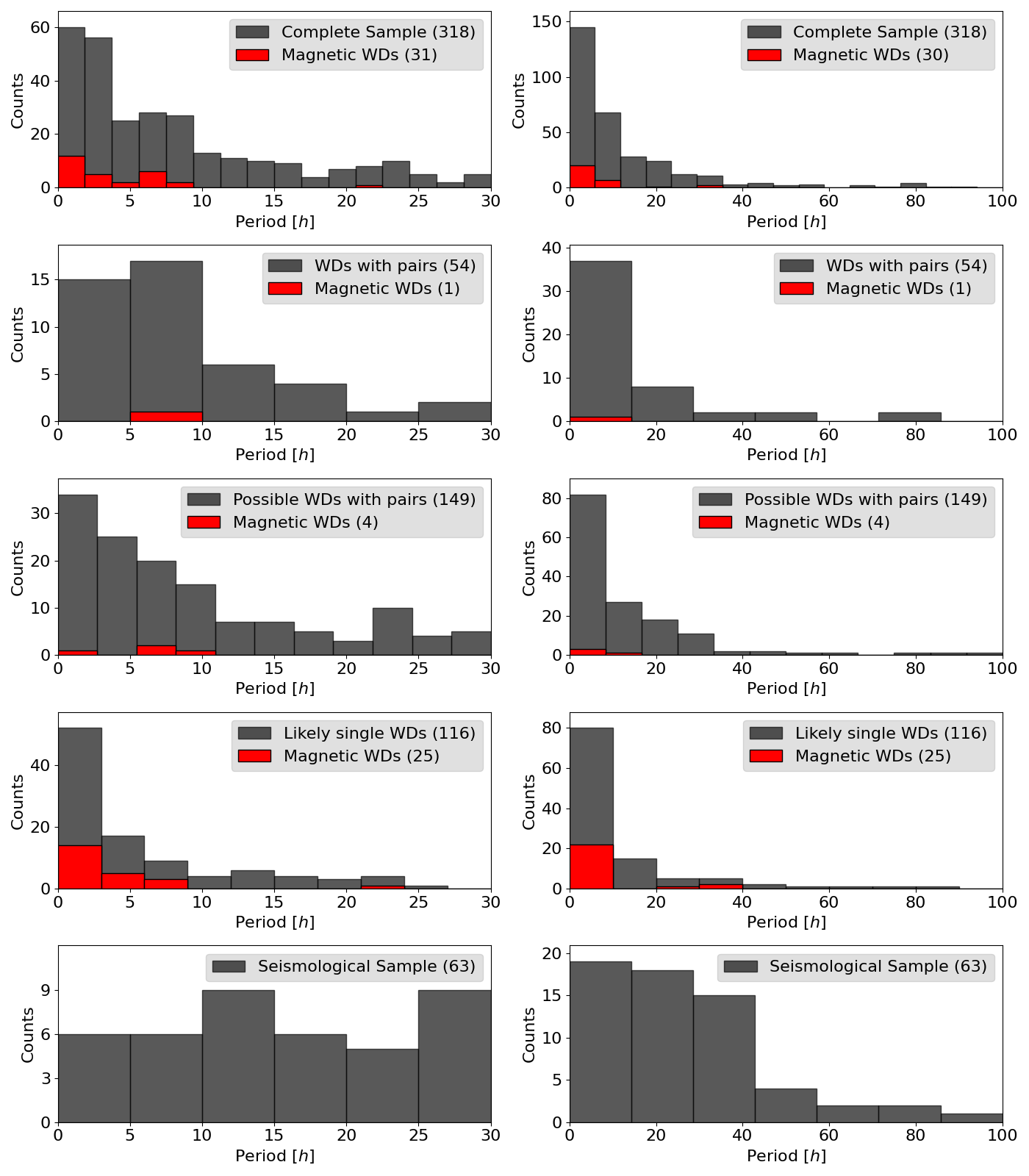}
    \caption{Variability period histograms for different samples of this work over and, last line, histograms of rotation periods from seismology. On the left, we present the histograms of periods up to $30$~h, and on the right, the histograms of periods up to $100$~h. The red bins show the magnetic population in our samples of WDs. The labels indicate the samples plotted as well as their number.}
    \label{rothist2}
\end{figure*}

\par Figure \ref{rothist2} shows the histograms of periods up to 100~h (right) and up to 30~h (left) for each sample listed in Table \ref{samples}. Each line of this figure displays the histograms of one sample, indicated by their legends. The last line of Figure \ref{rothist2} shows the same histograms but for the rotational periods determined by asteroseismology, which are listed in Table \ref{pulsatingtab}. The red bins represent the magnetic WDs in each sample.  

\par In the first line of Figure \ref{rothist2} we observe the distribution of our \textit{Complete Sample} with its population of 30 magnetic WDs indicated by red bins. 
The right panel shows that the variability periods are mostly concentrated in the shortest bins. Even considering only periods shorter than 30~$h$ (left panel) the two shortest bins encompass 36$\%$ of our sample. Moreover, this plot shows that our sample of magnetic WDs is concentrated in the shortest periods, presenting a median period of 3.4 hours.

\par The second line of Figure \ref{rothist2} shows the period histograms for the sample of \textit{WDs with close pairs}. This sample is composed of 54 stars, 1 of them with a detected magnetic field. The periods detected for this sample have a median of 8.1~hours. 
The third line of Figure \ref{results2} illustrates the period distributions for the sample of \textit{Potential WDs with pairs}. This sample is composed of 149 WDs, including 4 with magnetic field detected. Their median period is 7.6~h, close to the median period of the \textit{WDs with close pairs}.

\par Finally, the fourth line of Figure \ref{results2} shows the period distribution of the \textit{Likely single WDs} sample. This sample is composed of 115 stars, 25 of them with magnetic field detected. The right panel shows that the photometric rotation periods are very concentrated in the shortest period bin, 69~$\%$ of periods are shorter than 10~hours. The \textit{Likely single WDs} exhibit a median period of 3.9~$h$, very close to the median period exhibited by our complete magnetic sample (3.4~$h$). The 25 magnetic WDs included in the \textit{Likely single WDs} sample present an even shorter median period of 1.9~hours. 

\par The last line of Figure \ref{results2} displays the distribution of the rotational periods estimated through asteroseismology, compiled in Table \ref{pulsatingtab}. Despite the difference in sample sizes, this figure shows significant disparities between the rotational distributions derived from photometry and seismology. The seismological sample has a median rotational period of 24.3~$h$, which is over 6 times longer than the photometric rotational median period. Additionally, the distribution of rotational periods derived from seismology is broader compared to photometry, with the seismological and the photometric samples exhibiting a median absolute deviation of 13.9~$h$ and 3.5~$h$, respectively.

\par It is important to acknowledge that the rotation period calculated through asteroseismological models is an estimate of the mean internal rotational period of the star; while the photometric rotational period represents the rotation of the stellar surface. Moreover, the seismological sample and the \textit{Likely single WDs} sample are representative of different populations. The seismological sample is representative of the pulsating WDs, whose instability strips are very narrow in temperature range. In contrast, our \textit{Likely single WDs} is representative of the WDs with surface inhomogeneities, probably mostly caused by magnetic fields.

\begin{figure}[h]
    \centering
    \includegraphics[width = 0.47\textwidth]{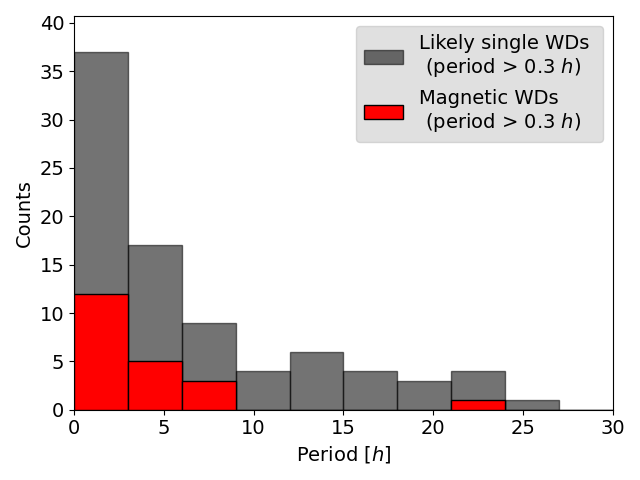}
    \caption{Rotational period distribution up to 30~h for the \textit{Likely single WDs} excluding periods shorter than 0.3~h. The red bins indicate the magnetic stars of the sample.}
    \label{rothist1}
\end{figure}

\par As explained in Section~\ref{data}, we excluded pulsation periods. However, to eliminate any doubts about the short periods in our \textit{Likely single WDs} sample, Figure~\ref{rothist1} shows the histogram of periods longer than 0.3~h. It is important to note that previous work has already detected periods shorter than 0.3~h for massive white dwarfs \citep[e.g.][]{Caiazzo21, Barstow95}. The inset plot confirms that even with a conservative approach of excluding periods shorter than 0.3~h, the majority of our sample still exhibits rotation periods shorter than 10~h. 

\subsection{Comparison with Ground-based observations}

\par
\begin{figure}[h]
    \centering
    \includegraphics[width = 0.5\textwidth]{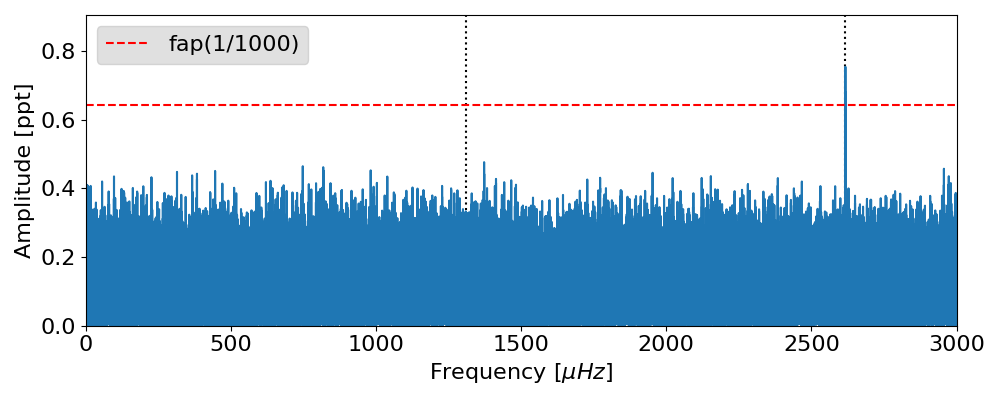}
    \caption{Fourier Transform of \textrm{TESS} data of star TIC~262548040. The horizontal red line indicates the false alarm probability FAP=1/1000 and the vertical dotted lines indicate the peak of 0.106~h and it's double. }
    \label{TIC262548040}
\end{figure}

 \citet{2023MNRAS.523.5598M} obtained time-resolved spectroscopy for two magnetic WDs in our sample at the Gemini Observatory. Based on the shifting positions of the Zeeman-split $H_{\alpha}$ components, they inferred a rotation period of 0.648~h for the star TIC~392797216 (LHS~2273). This star is included in our sample of \textit{Likely single WDs}, and the period reported is very close to the 0.68~h period we found in the \textrm{TESS} data. Using the same method, they found that the star TIC~262548040 (LHS~1243) has a rotation period of 0.216~h. Due to its low mass of 0.41~$M_{\odot}$, we classified this star as \textit{Possible WD with pair}. Figure~\ref{TIC262548040} presents the FT of the data obtained by \textrm{TESS} for this star, showing that only one peak at $2617\,\mu$Hz (0.106~h) is above the detection limit (horizontal dashed red line). The vertical dotted lines indicate the peak of 0.106~h and its double (0.212~h), showing that there is no significant peak at 0.212~h in the \textrm{TESS} data. Therefore, it shows that, in this case, the signal we detect in the \textrm{TESS} data is the harmonic of the real rotation period.
Despite the harmonic issue, \citet{2023MNRAS.523.5598M} confirms the rotation rate for two of the fastest rotators in our sample. 

\subsection{Pulsating White Dwarfs}
\label{results2}

\par From the 63 stars listed in Table \ref{pulsatingtab}, 45 were observed by \textrm{TESS}. We analyzed all of them, searching for nonpulsating periods. 
Most pulsating WDs do not show any frequency with amplitude above the detection limit, other than pulsation periods, or show only low amplitude at frequencies smaller than 4.6~$\mu$Hz (10 times the frequency resolution of a sector) in the FT. Some pulsating WDs exhibit dozens of pulsation frequencies in their FTs, leading to linear combinations at low frequencies. 
We only identified reliable variability periods outside the pulsation range (periods longer than 1500~s) for three of these stars: NGC~1501, TIC~7675859, and G226-29. Their \textsc{TESS\_localize} output parameters are compiled in Table \ref{tab3} and are described in the following subsections: \ref{ngc}, \ref{alestar}, and \ref{sg22629}.

\begin{table*}
    \begin{tabular}{|c|c|c|c|c|c|c|} \hline
    Name & TIC & FT Period [h] & Amplitude [$\sigma$] & Like & $\langle$ Height [$\sigma$] $\rangle$ & $Q/S$ \\ \hline 
NGC 1501 & 84306468  & $138 \pm 32$ & 15.377 & 1.0 & 7.46 & 1/1 \\ \hline
         &           & $39 \pm 2$ & 18.104 & 1.0 & 18.4 & 1/1 \\ \hline
         &           & $4.11 \pm 0.03$ & 14.571 & 1.0 & 15.4 & 1/1 \\ \hline
TIC 7675859  & 7675859  & $12.344 \pm 0.008$ & 8.768 & 1.0 & 8.1 & 5/6 \\ \hline
G226-29 & 199666369  & 0.93 - 3.25 & 4.129 - 13.934 & 1.0 & 9.4 & 27/29 \\ \hline

    \end{tabular}
    \caption{Non-pulsating periods from pulsating white dwarfs.}
    \label{tab3}
\end{table*}

\subsubsection{NGC~1501\label{ngc}}

\begin{figure}[h!]
    \centering
    \includegraphics[width = 0.5\textwidth]{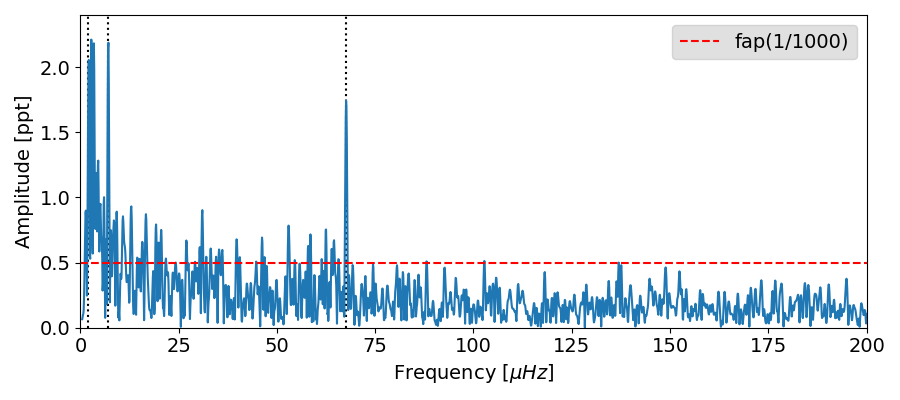}
    \caption{Fourier transform of \textit{NGC~1501} data at low frequencies.}
   \label{fig:ngc}
\end{figure}
\par The pulsating pre-white dwarf NGC~1501 was discovered to pulsate with periods from 5235~s to 1154~s, 
by \cite{Bond96}. They estimate a rotation period of 1.17~days from a mean splitting of $9.9\,\mu$Hz.
This star was observed by \textrm{TESS} during sector 19 for 25 days with 120~s exposures. We analyze its FT and detect three variability frequencies outside its pulsating frequency range. \textsc{TESS\_localize} confirmed that the source of these signals is NGC~1501. Figure \ref{fig:ngc} shows this FT up to $200\, \mu$Hz, where the three non-pulsating frequencies that we have detected are indicated by the vertical dotted lines and correspond to 4.11, 39, and 138~hours. None of these periods is consistent with the rotation period estimated by \cite{Bond96}.
Moreover, \textit{NGC~1501} is a planetary nebula covered by an ionized gas disk. Therefore, considering the complexity of this system, we are unable to interpret these periods. 

\subsubsection{TIC 7675859\label{alestar}}

\begin{figure}[h]
    \centering
    \includegraphics[width = 0.5\textwidth]{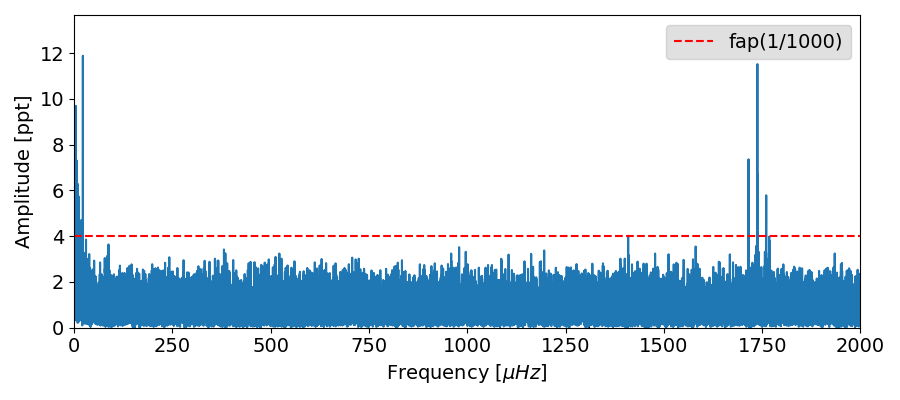}
    \caption{Fourier transform of \textit{TIC 7675859} data from sectors 52 to 54. }
   \label{TIC7675859}
\end{figure}
\citet{Romero24} determined the asteroseismological rotation period of 7.15~h for the DAV \textit{TIC~7675859} based on data from sectors 25, 26,40 52, 53, and 54. We observed the three components of a high-amplitude triplet in the data of sectors 52 to 54 (see Figure \ref{TIC7675859}). The triplet at $1738\, \mu$Hz is perfectly symmetric with a frequency spacing $\Delta \nu=22.47\,\mu$Hz.
Moreover, we also observed a significant peak at $22.44\,\mu$Hz (corresponding to 12.3~h), which is a very close value to the mean frequency spacing of this triplet. The peak at $22.44\,\mu$Hz is higher than all components of the triplet at $1738\,\mu$Hz, however, we cannot eliminate the possibility that it is a linear combination \citep{Kurtz15}. This peak of 12.3~h can also be a subharmonic of the real rotation rate or even the real rotation rate of the stellar surface itself, indicating differential rotation.

\subsubsection{G226-29\label{sg22629}}

\begin{figure}[h]
    \centering
    \includegraphics[width = 0.48\textwidth]{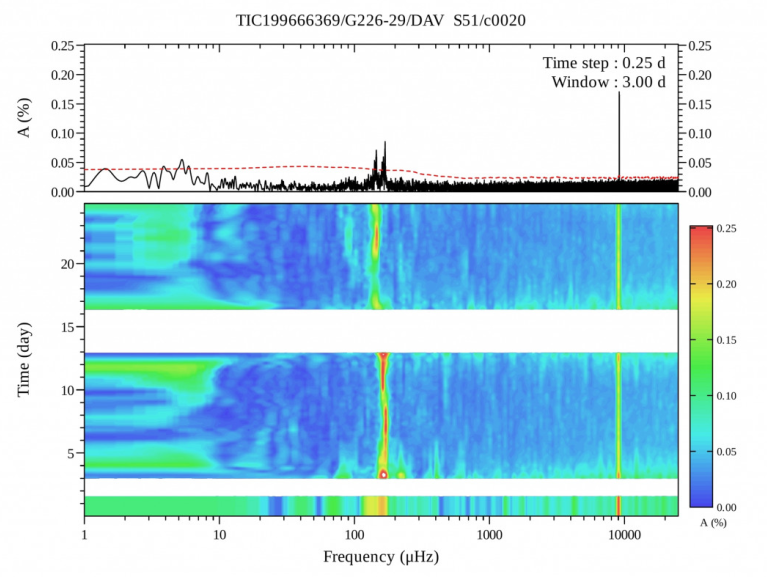}
    \caption{Running Fourier transform of G226-29 20~s data from sector 51. The tall peak on the right is the pulsation triplet around 109~s, while the region around 200~$\mu$Hz corresponds to the peaks around 4500~s.}
   \label{sft}
\end{figure}

\par
\begin{figure}[h]
    \centering
    \includegraphics[width = 0.48\textwidth]{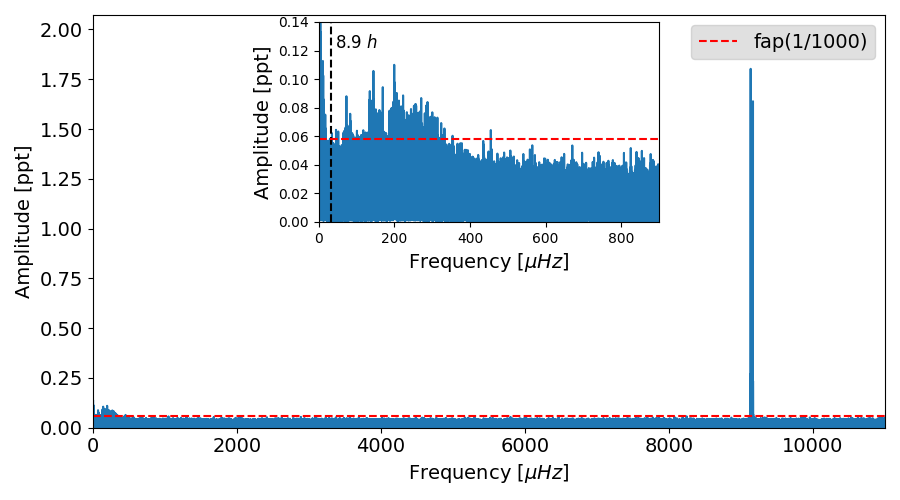}
    \caption{Fourier transform of the 20~s data for G226-29 from sectors 40 to 41 and 47 to 60. The inset plot zooms the FT at low frequencies. The red dashed line indicates the false alarm probability. The large peak on the right is the pulsation triplet at 109.47235s@1.81mma, 109.27929s@0.57mma, 109.08690s@1.62mma, while the region around 200~$\mu$Hz corresponds to the wandering peak around 4500~s.}
   \label{g22629}
\end{figure}

 G226-29 is the brightest known DAV, with a known pulsation triplet around 109~s \citep{Kepler83}. When studying this triplet, \citet{Kepler95} estimated the rotation period of the WD as 8.9~h. 
G226-29 was observed by \textrm{TESS} with 120~s exposures from sector~14 to 26, and with 20~s exposures from sectors 40 to 41 and 47 to 60. In total, it was observed along 29 sectors. Data from all sectors show a S/N$\geq 4.1$ wandering peak from 0.93~h ($297.3\,\mu$Hz) to 3.25~h ($85.5\,\mu$Hz). As an example, Figure~\ref{sft} plots the Running Fourier Transform of sector~51 data. This figure shows the known stable triplet around $\sim 9200\,\mu$Hz and a peak that wanders from $\sim 145\,\mu$Hz to $\sim 168\,\mu$Hz. 
Figure~\ref{g22629} shows the complete FT of all 20~s data combined, where the wandering peak appears as a broad band at low frequencies (see the inset plot).   

\par The methodology described in Section~\ref{localize} assumes that the variability signals are stable. Therefore, to verify whether this wandering peak comes from G226-29 we had to apply the \textsc{TESS\_localize} test differently. For each sector, we obtain the highest peak exhibited in such a FT range and run the \textsc{TESS\_localize} code for this period. As detailed in Table~\ref{tab3}, \textsc{TESS\_localize} confirmed that this wandering peak indeed comes from G226-29; however, this variability remains a mystery.

\par The rotational period from seismology is indicated by a vertical black dashed line in Figure~\ref{g22629}, showing the signal we are detecting is not consistent with the rotation of the stellar inner. On the other hand, this signal is very long to be consistent with pulsations. This leads us to suspect differential rotation; that is, the wandering peak could be a manifestation of rotation of the stellar surface, that is rotating faster than the inner. Although the star has an upper limit on the magnetic field around 10~kG \citep{Valyavin06}, subsurface magnetic fields, like those in Section~\ref{our-models}, are a possibility. Only radial differential rotation does not explain why this peak varies so much, but an angular differential rotation, like the sun  \cite[see e.g.][]{1990ApJ...351..309S, 2000SoPh..191...47B}, could.

\section{Evolutionary models \label{tayno}}

\par If the rotation periods we detect for the \textit{Likely single WDs} sample are due to surface inhomogeneities caused by magnetic fields, we need to calculate models that generate magnetic fields and reproduce the order of magnitude of the rotation rates we are observing. \cite{Bagnulo20,2023ApJ...944...56A} show that at least 20\% of all white dwarfs are magnetic, and \cite{Bagnulo22,CastroTapia24a,CastroTapia24b,Blatman24a,Blatman24b} show dynamo induced by phase separation during crystalization cannot be the origin for most magnetic white dwarfs.
When we study the rotation rates of evolved white dwarf stars, 
one question that arises is how the evolution changes the internal velocities and how much of the initial angular momentum is lost during the expressive mass loss phase of AGB before the white dwarf phase.
The models included in this section have internal magnetic fields generated by dynamos caused by the rotation of the star and angular momentum transfer. We calculate their rotation rates, internal and at the photosphere, to compare with the observed ones and observe how they change --- or not --- with the effective temperature of the star.

\subsection{Observational and modeling context} 

\par Using ground-based photometric observation time series from mean sequence stars spanning late-F to early-M spectral types, \citet{Fritzewski2021} determined 279 rotation periods. They found that the rotation periods range from 0.5~d to 32~d, with trends of slow-rotating FGK stars and fast-rotating K dwarfs.
In another recent investigation, \cite{Labadie-Bartz2023} explored the rotational characteristics of chemically peculiar (CP) stars using TESS data. Among their findings, they found that the rotation period distribution of CP stars is centered around 3 days. CP stars are observed within the spectral types from early B to early F.
Since all stars with spectral type roughly from A to K will become WDs, both works evidence the rotational period of WD's progenitors is of the order of days. We have shown in this work that the median photometric rotation period of WDs is 3.9~h; therefore, these findings suggest that a robust mechanism of internal angular momentum transfer operates during the evolution of these stars.

The asteroseismology of low and intermediate mass stars enabled the measurement of the internal rotation rate of WD progenitors at different evolutionary phases \citep[e.g.,][]{Mosser2012, Deheuvels2012, Deheuvels2014, Mosser2012, Deheuvels2015, DiMauro2016, DiMauro2018, Gehan2018, Tayar2019, Deheuvels2020}.
These inferred rotation rates have been used to constrain models of angular momentum transfer processes.
\citet{AM-I} and \citet{AM-V} found that the transport of angular momentum must be more efficient in more massive stars during the subgiant and red giant phase, even if the efficiency of the internal transport of angular momentum decreases with the evolution of the star. They suggest that the physical nature of the additional mechanism may be different in subgiant and red giant stars.
\citet{Eggenberger2022} find that the core rotation rates of the red giant branch models are nearly insensitive to the initial rotation velocity.

\par On theoretical grounds, \citet{Fuller2019} deduced a new expression for the Tayler instability that leads to larger magnetic field amplitudes, more efficient angular momentum transport, and smaller shear than predicted by the original Tayler–Spruit dynamo.
\citet{AM-II} applied the new prescription for subgiant and red giant models and found that it leads to low core rotation rates after the main sequence that are in better global agreement with asteroseismic measurements than those predicted by the original Tayler-Spruit dynamo. However, it fails simultaneously to reproduce the asteroseismic measurements available for subgiant and red giant stars.

\citet{Hartogh2019} found that the rotational period of the models with additional viscosity is too large in the white dwarf phase. However, the rotational periods match white dwarf rotation periods of the order of 1~day when they exclude the additional effect during the core-helium-burning phase.
Therefore, they proposed that the efficiency of the unknown angular-momentum transfer process must decrease during the core-helium burning phase. Applying the \citet{Fuller2019} formalism, \citet{AM-III} found that, when the dynamo effect is turned off at the end of the core helium burning phase, the WD rotation rates decrease to the order of days.
They concluded that the \citet{Fuller2019} dynamo formalism cannot be the sole solution to the missing process of angular momentum transport in intermediate-mass stars.

\par Magnetic fields are present in low-mass stars and, when strong enough, can change the rotation by exerting a torque capable of reducing the differential rotation or even imposing uniform rotation throughout the star. \citet{Moyano2023} showed that the transport of angular momentum in radiative zones during the main sequence of low-mass stars must be efficient. They suggested that the internal magnetic fields are a strong candidate for the missing physical ingredient in stellar interiors. In addition, they point out that at least one efficient process able to neutralize the development of differential rotation in stellar interiors must act during the whole evolution of these stars.

\subsection{Our models} \label{our-models}

We computed evolutionary sequences using the Modules for Experiments in Stellar Astrophysics code
\citep[\textsc{MESA,}][]{Paxton2011, Paxton2013, Paxton2015, Paxton2018, Paxton2019, Jermyn2023}, release 22.11.1.
Our models have an initial mass of
$M_{\text{i}} = 1.5~M_{\odot}$
\citep[following][]{Hartogh2019}
and an initial metallicity of
$Z = 0.02$, 0.002 and 0.001.
The initial mass was chosen to be above the lower limit of
$1.3~M_\odot$ for the convective core in the main sequence
\cite[e.g.][]{Aerts2023arXiv231108453A},
leading to a WD with a mass close to the mean mass of most white dwarf stars \citep[e.g.][]{kepler07,2023MNRAS.518.3055O}.
The models are computed from the zero-age main sequence (ZAMS) until a white dwarf cooling effective temperature of 8500~K.
Our models start with equatorial surface rotation velocities of 10 and 50~km~s$^{-1}$, typical of main-sequence stars with 1 and 2 solar masses.
These values are assumed to be the solid-body rotations at the ZAMS,
which is a reasonable and common assumption for this type of star
\citep[see, e.g.,][]{Bouvier1997,Granada2014,Amard2019,Deal2020,Nguyen2022,Douglas2024}.
For further information on the models, see \nameref{AppendixC}.

The mixing of elements and the transport of angular momentum due to rotation are implemented in
\textsc{MESA}
following closely
\citet{Heger2000,Heger2005}.
Two efficiency factors must be set to calibrate the diffusion coefficients: the contribution of rotationally induced instabilities to the diffusion coefficient is reduced by the factor $f_{\text{c}}=1/30$, and the sensitivity of rotationally induced mixing is $f_{\mu}=0.05$.

We consider the following angular momentum transport mechanisms:
dynamical shear instability,
Solberg-Hoiland,
secular shear instability,
Eddington-Sweet circulation,
Goldreich-Schubert-Fricke and the
Tayler-Spruit dynamo.

The final masses of our models are
0.546~$M_{\odot}$ for $Z=$~0.02,
0.577~$M_{\odot}$ for $Z=$~0.002,
and 0.585~$M_{\odot}$ for $Z=$~0.001.
This aligns with the expectations that stars with higher metallicity should form less massive CO cores and lose more mass in the AGB phase than stars with the same initial mass but lower metallicity
\citep[e.g.,][]{Dominguez1999,Catalan2008,Romero2015,Choi2016}.
Similar models with initial masses of 2.5~$M_{\odot}$ reach the white dwarf phase with masses around $0.6~M_\odot$.

\begin{figure}
    \centering
    \includegraphics[width = 0.5\textwidth]{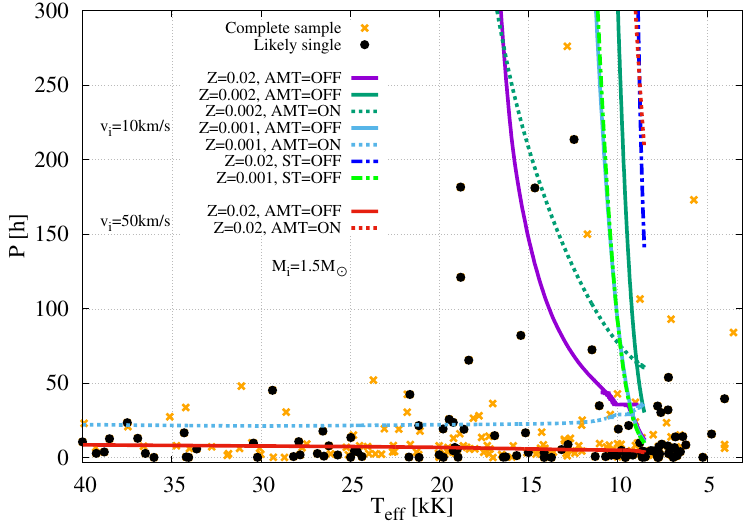}
    \caption{Observational data of the surface rotation of white dwarf stars (orange crosses for the complete sample and black dots for the likely single sample}) are compared against evolutionary models showing the rotation period in the white dwarf cooling track. We compare models with the six mechanisms of angular momentum transport turned off (solid lines) and on (dotted lines) during the whole evolution. For an initial rotational velocity of $v_{\rm i}=$~10~km~s$^{-1}$, we present models with metallicities of $Z=$~0.02 (purple), 0.002 (dark green) and 0.001 (light blue). We also compare two models where the Tayler-Spruit dynamo is the only mechanism that is turned off during the computations (dark blue and light green dotted-dashed lines). For an initial rotational velocity of $v_{\rm i}=$~50~km~s$^{-1}$, we compare two models with $Z=$~0.02 (red lines). All model sequences have an initial mass of 1.5~M$_{\odot}$.
    \label{fig:Teff-P+data}
\end{figure}

In Figure~\ref{fig:Teff-P+data}
we show the evolution of the surface rotational period ($P$) versus the effective temperature ($T_{\rm{eff}}$) during the cooling track for a set of model sequences with different metallicity and initial rotation velocity.
The surface rotational period is computed via $P=2 \pi r/v$, where $r$ is the radius of the model and $v$ is the rotational velocity at the equator.
Sequences that consider the six mechanisms of angular momentum transport are shown as dotted lines (AMT = ON), and solid lines represent sequences with all mechanisms turned off (AMT=OFF).

For $v_{\rm i} =$~10~km s$^{-1}$ and $Z =$~0.02 (purple line in
Figure~\ref{fig:Teff-P+data}), we find that the white dwarf models with AMT = OFF decrease the average equatorial rotation period on the surface from $P=$~300~h to $\sim$~30~h when cooling from $\sim$~17~kK to 8500~K. The sequence with AMT=ON does not even reach 300~h before cooling down to the same temperature (and therefore does not appear in the plot).

For sequences with $Z=$~0.002 (dark green lines in
Figure~\ref{fig:Teff-P+data}), we find that if AMT=OFF (solid line), the models brake from $P=$~300~h to 40~h only at the very end of the cooling track ($T_{\rm eff}<$~10~kK), while the AMT=ON sequence (dotted line) has already spun down to 300~h when $T_{\rm eff}<$~17~kK, although it reaches only $P \sim$~60~h at $T_{\rm eff}=$~8500~K.

The internal redistribution of the angular momentum is very efficient early in the evolution of low-mass stars
\citep[e.g.,][]{Mosser2012}.
This suggests that when AMT=ON, a large amount of mass lost by winds during the AGB phase carries away most of the angular momentum, and thus the surfaces of the models rotate slower in the WD phase.

The difference in turning the angular momentum transport mechanisms on and off is much more pronounced for low metallicity.
($Z=0.001$, light blue in
Figure~\ref{fig:Teff-P+data}).
While AMT=OFF (solid line) follows the same trend as for higher metallicities (that is, it only reaches low $P$ at the end of the cooling track), the sequence with AMT=ON (dotted line) cools from $T_{\rm eff}=$~40~kK down to 8500~K with approximately constant rotation period of $P=$~25~h, closer to most of the observed period distribution.

In Figure~\ref{fig:Teff-P+data}
we also show two sequences of models with $v_{\rm i} =$~10~km~s$^{-1}$
where the Tayler-Spruit mechanism was turned off, and all the other five mechanisms were kept on throughout evolution (ST = OFF, dotted-dashed lines).
For $Z=$~0.001 (light green line), the result is similar to that of the same metallicity, but with all six angular momentum transport mechanisms turned off. This suggests that the Tayler-Spruit mechanism dominates over the other mechanisms from the point of view of rotation analysis during the WD phase.

Finally, we present two sequences with $v_{\rm i} =$~50~km~s$^{-1}$ and $Z=$~0.02.
These models show that observational data with a shorter rotation period ($P<$~50~h) are better adjusted by models with AMT=OFF,
while models with AMT=ON present a rotation period shorter than 300~h only at the end of white dwarf cooling when $T_{\rm eff}<$~10~kK.
This suggests that if we disregard the Tayler-Spruit mechanism throughout the evolution, models with a higher initial velocity (50~km~s$^{-1}$) fit the white dwarf rotation data better if the initial mass is 1.5~M$_{\odot}$.

\par The dominant factor for the spin-up of the external layers during the white dwarf cooling phases are the He-convection zones that arise at $T_\mathrm{eff}\simeq 32\,000$~K, followed by the H-convection zone if the star has a H envelope, around $T_\mathrm{eff}\simeq 14\,000$~K. The movement of mass in the convection zones efficiently redistributes the angular momentum.

\par
\begin{figure}
    \centering
    \includegraphics[width =0.5\textwidth]{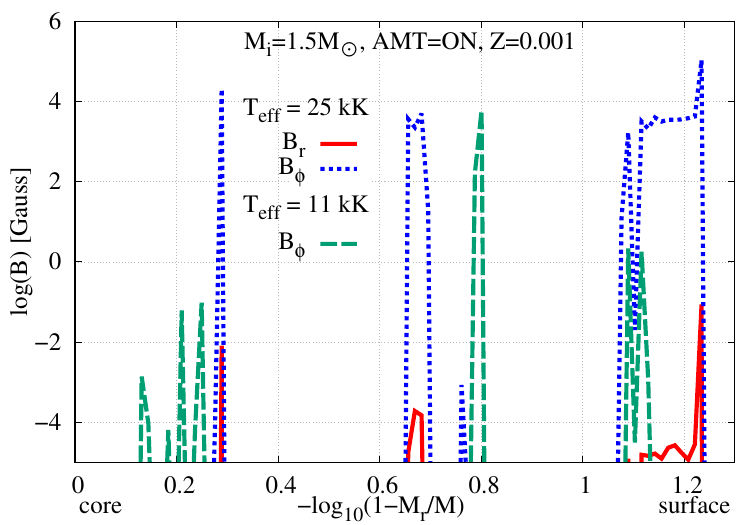}
    \caption{ Internal magnetic field at $T_\mathrm{eff}=25\,000$~K and 11\,000~K, generated by the Tayler-Spruit dynamo during the evolution due to differential rotation caused by angular momentum transfer throughout the stellar evolution. $B_r$ is the radial field and $B_{\phi}$ the azimuthal one. The properties of the initial model in the ZAMS are $M_{i}=1.5~M_{\odot}$, $Z=0.001$, and $v_{i}=10$~km~s$^{-1}$. For $T_{\rm eff}=25\,000$~K, the solid red line shows the radial component and the dotted blue line represents the azimuthal one. For $T_{\rm eff}=11\,000$~K, the dashed green line shows the azimuthal component, and the radial component does not have enough intensity to appear in the plot. The six angular momentum transfer mechanisms operate throughout evolution. Magnetic field values are negligible beyond 1.3 on the $x$-axis, i.e., towards the surface. \label{magnetic}}
\end{figure}

 Also, the models with AMT=ON generate internal magnetic fields due to differential rotation, as shown in Figure~\ref{magnetic} for a model sequence with $M_{i}=1.5~M_{\odot}$, $Z=0.001$, and $v_{i}=10$~km~s$^{-1}$ (corresponding to the dotted light blue line in Figure~\ref{fig:Teff-P+data}).
According to \citet{Spruit2002}, the Maxwell stresses transport angular momentum in the radial direction at a rate proportional to the product $B_{r}B_{\phi}$.
In Figure~\ref{magnetic}, the $x$ axis represents the mass coordinate of the model, with the core on the left and the surface on the right, and the $y$ axis shows the intensity of the magnetic field components $B_{r}$ and $B_{\phi}$ in the internal part of the models at two different temperatures during the cooling track. The chosen temperatures, 25\,000~K and 11\,000~K, are representative of the DAV and DBV variable classes.

\par Figure~\ref{magnetic} shows that, during cooling, the azimuthal component of the magnetic field is always greater than the radial component.
In some regions, the magnetic field reaches an intensity of $\sim 10^4$~G.
We also found that both components tend to be less intense as the models cool, and the outer $\sim 5\%$ (by mass) of the models show no significant magnetic fields.

As a general behavior, the models suggest the following.
During cooling, the Tayler-Spruit dynamo mechanism acts smoothly in the core, flattening the rotation profile up to the He layer.
On the other hand, a convective zone appears at the
base of the He envelope around 32\,000~K
and increases in size inward as the models cool, causing angular momentum to be transferred from the core and the He zone.

Around 14\,000~K a convective zone develops at the base of the H envelope,
and then the rotational speed on the surface is as high as that of the fastest part of the core and the He zone.
The exact temperature for this to occur depends on the metallicity and, especially, on the history of mass loss at the end of the AGB phase.
This is because the last thermal pulses directly interfere with the amount of H remaining in the model that enters the cooling track, modifying the convection in this zone.

Thus, the internal rotation profile of the models during cooling depends, to some extent, on angular momentum transfer mechanisms, but also strongly depends on convection in the He and H zones.

\section{Conclusion \label{conclusion}}
\par We have reported significant variability periods for 318 white dwarfs, as well as their variability amplitude and quality criteria. We divided the 318 stars composing the \textit{Complete Sample} into three sub-samples. 54 of them were classified as \textit{WDs with close pairs}, and their variability periods can be due to orbital motion or rotation. The other 149 were classified as \textit{Potencial WDs with pairs} due to their low masses and/or high temperatures \citep[e.g.][]{Steen24}. Finally, we classified 115 as \textit{Likely single WDs} and therefore attributed their variabilities to rotation.

\par Our sample of 30 magnetic WDs are massive, fast rotators, and cold; while our sample of 54 \textit{WDs with close pairs} are well distributed on the mass range and presents a weak tendency of longer periods to massive WDs.

\par Our \textit{Complete Sample} of 318 WDs exhibits a temperature distribution expected for DA WDs, peaking around $T_\mathrm{eff}\simeq 12,000~$K. On the other hand, its mass distribution is biased towards low-mass stars because it is a magnitude-limited sample. The mass distribution corrected by $1/V_{max}$ presents, in density, a lack of WDs with a mass around 0.6~$M_{\odot}$ and an excess of massive WDs, which are dominated by our little sample of 30 magnetic WDs.
These peculiarities are not present in the corrected mass distribution of our initial sample of 845 stars, which leads us to conclude that these characteristics were introduced by the \textsc{TESS\_localize} high-quality selection. In this second selection, 79~$\%$ of the stars discarded by \textsc{TESS\_localize} was done because the software was unable to perform a high-quality fit.
Therefore, it is possible that \textsc{TESS\_localize} have selected the highest amplitude variabilities and, consequently, the WDs with the strongest magnetic field. It could lead to a mass distribution dominated by magnetic WDs resulting from binary interactions and possibly explain the lack of 0.6~$M_{\odot}$ WD density.

\par The stellar density of our sample is completely dominated by the \textit{Likely single WDs}, which, as already discussed, is probably composed of magnetic WDs. A piece of evidence is that \textsc{TESS\_localize} reduced our initial sample by 62$\%$; however, it reduced only by 11.8$\%$ our initial sample of 34 known magnetic WDs. 

\par The \textit{Likely single WDs} present a median rotational period of 3.9 hours that is far shorter than the median rotational period from seismology estimates, 24.3 hours. However, this is an expected result, once both estimates of rotation represent different populations. The seismological sample represents the mean internal rotation of pulsating WDs, while our \textit{Likely single WDs} represents the rotation of the stellar surface of WDs with inhomogeneities at their surface.

\par From the 63 WDs with rotation period determined by asteroseismology, we detected nonpulsating periods in the \textrm{TESS} data for 3 of them: NGC~1501, TIC~7675859, and G~226-29. We were not able to interpret with certainty any of these periods. This failure in finding photometric rotational periods for the seismological sample using \textrm{TESS} data reinforces that the seismological and the \textit{Likely single WDs} samples represent distinct populations. Up to the moment, we do not know any pulsating WD with magnetic field confirmed.

\par Using ground-based observations, \citet{2023MNRAS.523.5598M} confirmed the rotation period we detected in the \textrm{TESS} data for two of the fastest rotators in our sample; however, they also provided an example of half-period aliases. It attempts to the fact that we may be detecting and reporting half of the real rotation period for other stars. On the other hand, this work confirmed rotation periods of 0.68~h and 0.21~h for two magnetic WDs in our sample; therefore, it does not invalidate our findings. 

\par For 5 of the 318 WDs in our sample, all of them included in our sample of \textit{Likely single WDs}, the variabilities detected in the \textrm{TESS} data were also reported by \cite{farihi2023nearby}. They are TIC~328029653, TIC~251080865, TIC~251903434, TIC~204440456 and TIC~321979116. The periods reported agree with our findings. They claim that the variability found in the WD~2138-332 data (TIC~204440456) is the stellar rotation period, almost certainly due to magnetism. In addition, the orbital period of the EMLV TIC~308292831 was already reported by \citet{2021ApJ...922..220L}.

\par Figure \ref{fig:Teff-P+data} presents two models that at 40~kK reach rotation periods consistent with the observational findings of this work.  We find that periods shorter than 50~h are well-fitted by models with $Z=$~0.02, $v_{\rm i}=$~50~km~s$^{-1}$ and angular momentum transport set to OFF, or by models with $Z=$~0.001, $v_{\rm i}=$~10~km~s$^{-1}$ and angular momentum transport set to ON. 
A smaller part of the observational data presents a large dispersion of the rotation period and is concentrated at $T_{\rm eff} <$~20,000~K. Our remaining models best fit these stars. 

\par While other studies have shown that the internal transfer of angular momentum should be intense in the early phase of the evolution of low-mass stars, our study confirms that if we seek to match most of the white dwarf rotational data, this redistribution should cease or be less intense before the end of the AGB phase, when stars lose the highest amount of mass and, therefore, angular momentum.
If we assume that stars with lower metallicity will have a lower mass loss rate due to winds at the top of the AGB phase \citep[e.g., see discussion in][]{Hofner2018}, this means that they will lose less angular momentum and therefore rotate faster in the WD phase.
However, when we consider that more angular momentum transfer mechanisms have been activated since the beginning of the evolution, more angular momentum will be transferred to the surface and lost during the AGB phase.
Thus, in terms of the WD rotation period, lowering the metallicity is equivalent to reducing the angular momentum transfer mechanisms in the early evolution.
Due to their intrinsic faintness, the observed white dwarfs are mainly nearby galactic disk objects, most likely with metallicity closer to the Sun's. 

\par In general, our models agree with the most recent results in the literature in the sense that different initial parameters can be used to create models that are in agreement with a particular part of the observational data, but there is a degeneration in the parameter space (i.e., the initial velocity of rotation, initial mass, metallicity, etc.) that could be better resolved if we knew more precisely how the redistribution of angular momentum occurs in stellar interiors. 

\par In the future, we will analyze the \textrm{TESS} data observed after October 2023, calculate models with the Fuller dynamo prescription, and extend the mass range of the models, as the data include white dwarfs with a range of masses. 
We intend to perform spectropolarimetry to check if our \textit{Likely single WDs} are magnetic.
We also intend to confirm the rotational periods using high-resolution spectroscopic observations to detect the NLTE line core rotational broadening, as in \citet{2005A&A...444..565B}.

We thank the anonymous referee for his valuable comments and suggestions, that improved our paper significantly.
This work was carried out with the financial support of the Conselho Nacional de Desenvolvimento Científico e Tecnológico (CNPq), and by the Coordenação de Aperfeiçoamento de Pessoal de Nível Superior - Brasil (CAPES). Most of the calculations were performed at the Texas Advanced Computing Center, located at the University of Texas in Austin. We thank Stéphane Charpinet for maintaining the Evolved Compact Stars with TESS on TASOC and for providing the running Fourier Transform of the star G226-29. This research has made extensive use of NASA’s Astrophysics Data System Bibliographic Service (ADS), SIMBAD, and MAST. We used ASTROPY (https://www.astropy.org/), \textsc{TESS-LS} (https://github.com/ipelisoli/TESS-LS), \textsc{TESS\_localize} (https://github.com/Higgins00/TESS-Localizer),
\textsc{lightkurve} (https://docs.lightkurve.org/) and \textsc{PYRIOD} (https://github.com/keatonb/Pyriod).
\appendix

\section*{Appendix A - Known RR Lyrae Star} \label{AppendixA}
\begin{figure}[h!]
    \centering
    \includegraphics[width = 1.0\textwidth]{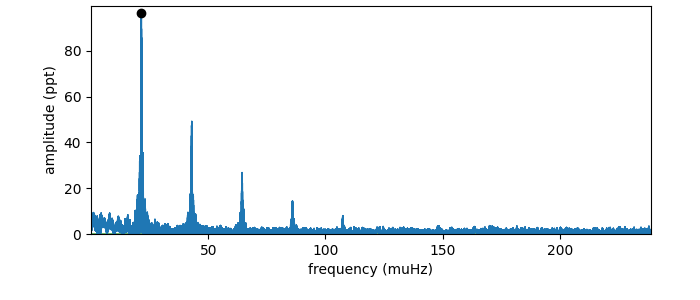}
    \caption{BPM~24754 light curve from sectors 12, 39, and 66.}
    \label{rrlyrae}
\end{figure}

\par BPM~24754 (TIC~367227831) is a known variable DAV white dwarf. Figure~\ref{rrlyrae} shows its FT of \textrm{TESS} data plotted using the Pyriod software \citep{2020AAS...23510606B}. In this case, we know that the primary signal at $21.49~ \mu$Hz (12.92~h) and its harmonics are the contamination from the known RR Lyrae variable \textit{Gaia DR3 5923100101270979328} (TIC 367227833). The results of software \textsc{TESS\_localize} for the three sectors where the star is observed (12, 39, and 66) are presented in Table \ref{tabA}.

\begin{table}[h!]
    \begin{tabular}{|c|c|c|c|c|c|}  \hline
    Sector & Source Gaia DR3& pvalue & Relative Likelihood & $\chi^2$ &  \textit{Height} [$\sigma$] \\ \hline
    $12$ & $5923100101270979328$ & $0.214$ & $0.75$ & $44.40$ & $42.40$ \\ \hline
    $12$ & $5923100101249048320$ & $0.071$ & $0.19$ & $44.40$ & $42.40$ \\ \hline
    $39$ & $5923100101270979328$ & $0.503$ & $0.83$ & $312.71$ & $17.30$ \\ \hline
    $39$ & $5923100101249048320$ & $0.191$ & $0.11$ & $312.71$ & $17.30$ \\ \hline
    $66$ & $5923100101270979328$ & $0.829$ & $0.96$ & $62.47$ & $37.22$ \\ \hline
    $66$ & $5923100101249048320$ & $0.198$ & $0.03$ & $62.47$ & $37.22$ \\ \hline
    \end{tabular}
    \caption{\textsc{TESS\_localize} results to the signal 12.92~h in present in the light curve of the DAV BPM~24754. }
    \label{tabA}
\end{table}

\textsc{TESS\_localize} estimates that the probability that the signal source is the RR Lyra star is $75\%$ in sector 12, even the parameters \textit{pvalue} and \textit{Height} indicating good values. The \textit{Relative Likelihood} of the RR Lyrae star being the source of the signal increases in the subsequent sectors, with the mean like reaching $85\%$. The second source is likely the same for all three sectors, \textit{Gaia DR3 5923100101249048320}. This object is located $3.06\arcsec$ away from the RR Lyrae star.

\section*{Appendix B - Inicial Sample of 845 WDs}
\label{AppendixB}
\par
\begin{figure}[ht!]
    \centering
    \includegraphics[width = 1\textwidth]{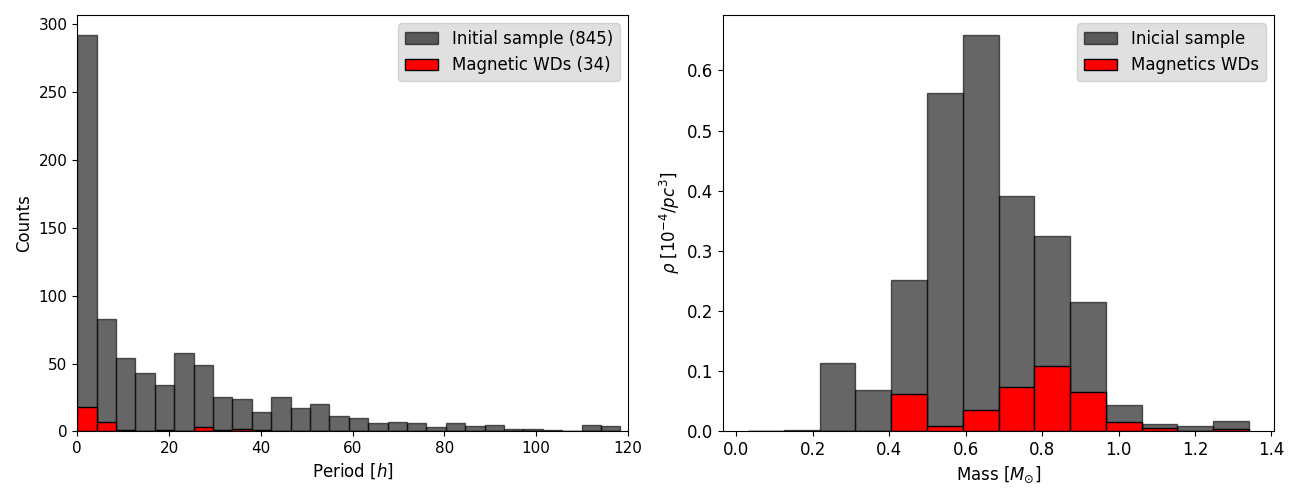}
    \caption{Mass distribution for our \textit{Initial} corrected by the $1/V_{max}$ volume.}
    \label{initialsample}
\end{figure}

 Our initial sample is composed of 845 WD stars exhibiting a non-pulsating period in their photometric data from \textrm{TESS} that, at least for one sector, \textsc{TESS\_localize} indicates the WD as the most probable signal source. As explained in Section \ref{localize}, most of our initial sample was discarded because \textsc{TESS\_localize} was unable to obtain a high-quality fit for the data. That means that for most of the stars in this sample we are not able to determine with high confidence whether the source of the signal we detected is, in fact, the WD. Figure \ref{initialsample} presents the mass distribution corrected by the $1/V_{max}$ method and the period distribution for our initial sample. The corrected mass distribution is centered around approximately 0.6~$M_{\odot}$, as expected for WD stars. Magnetic WDs are more massive, exhibiting a distribution around approximately 0.8~$M_{\odot}$. The period distribution is heavily concentrated in the shortest period bins, with the sample having a median period of 11.85~$h$ and a median absolute deviation of 11.56~$h$. 

\section*{Appendix C - Evolutionary Models Input}
\label{AppendixC}

Hydrogen and helium burning are computed using the
\texttt{pp\textunderscore and\textunderscore cno\textunderscore extras.net} network that accounts for 25 isotopes and 79 net reactions.
Convection is treated using the formulation of the mixing$-$length theory
\citep{Bohm1958MLT} in the variation of
\citet{Henyey1965MLT} allowing the convective efficiency to vary with the opacity.
We set the diffusion of elements for the whole net; each isotope in the network is treated as its own class.

Following the MIST project
\citep{Choi2016},
$\alpha_{\text{MLT}}=1.82$ is adopted as the mixing length parameter.
We consider the Ledoux criterion of stability,
which considers the influence of composition gradients on mixing.
Semiconvection is considered unstable regions by the Schwarzschild criterion but stable by Ledoux,
with an efficiency parameter
$\alpha_{\text{sc}}=0.1$.
Thermohaline mixing is included throughout evolution, with efficiency
$\alpha_{\text{th}}=666$.

Exponential overshooting is set to $f=0.008$ in the core and 0.0087 in the shell.
We handle mixing in the convective zones using the
convective premixing scheme.
For the boundary conditions of the atmosphere, we use the $T(\tau)$ Eddington relation with varying opacity for most of the evolution and hydrogen atmosphere tables for cool white dwarfs
\citep{Rohrmann2011}
when the models are on the cooling track and the effective temperature is below 10,000~K.
The mass loss by stellar winds is taken into account using the
\citet{Reimers1975}
scheme with
$\eta_{\text{R}}=0.5$
for the red giant branch phase and the
\citet{Bloecker1995}
scheme with
$\eta_{\text{B}}$
varying between 0.1 and 1 for the asymptotic giant branch phase.

\section*{Appendix D - Variabilities Periods Table}
\label{AppendixD}

\begin{longtable}[h]{|c|c|c|c|c|c|c|c|c|}
    \caption{Photometric rotation Periods.}\label{rottab}\\
    \hline
    TIC & Period [h] & {\footnotesize Amplitude [$\sigma$]} & Like & {\footnotesize $\langle$ Height [$\sigma$] $\rangle$} & T$_{\rm{eff}}$ [$K$] & Mass [$M_{\odot}$] & $Q/S$ & Info \\ \hline \endfirsthead
 
\hline TIC & Period [h] &  {\footnotesize Amplitude [$\sigma$]} & Like & {\footnotesize $\langle$ Height [$\sigma$] $\rangle$} & T$_{\rm{eff}}$ [$K$] & Mass [$M_{\odot}$] & & Info \\\hline \endhead
144002497	&	7.6213	$\pm$	0.0022	&	48.042	&	1.00	&	42.8	&	72910	&	0.606	&	2/2	&		\\	\hline
837076117	&	11.468	$\pm$	0.007	&	7.899	&	1.00	&	11.2	&	81335	&	0.65	&	1/2	&	DA	\\	\hline
307982318	&	1.53006	$\pm$	7e-05	&	4.803	&	1.00	&	5.0	&	31880	&	0.67	&	1/3	&	hotDAV	\\	\hline
88564975	&	8.9702	$\pm$	0.003	&	173.775	&	1.00	&	92.9	&	80000	&	0.67	&	3/3	&		\\	\hline
136884288	&	0.6053	$\pm$	0.0006	&	7.730	&	1.00	&	13.6	&	8440	&	0.81	&	1/1	&	DAHe	\\	\hline
383647517	&	13.924	$\pm$	0.014	&	13.263	&	1.00	&	6.2	&	6754	&	0.545	&	7/16	&	DQpec	\\	\hline
1102584065	&	26.187	$\pm$	0.022	&	17.910	&	1.00	&	9.1	&	110000	&	0.57	&	12/13	&	DO	\\	\hline
436747174	&	4.166	$\pm$	0.026	&	290.193	&	1.00	&	59.7	&	22500	&	0.532	&	1/1	&	DA-CV+M\\	\hline
434196824	&	3.904	$\pm$	0.023	&	53.066	&	1.00	&	62.4	&	14491	&	0.258	&	1/1	&		\\	\hline
345265261	&	3.759	$\pm$	0.022	&	7.601	&	1.00	&	12.2	&	38780	&	0.843	&	1/1	&	DA	\\	\hline
321272216	&	4.408	$\pm$	0.031	&	34.385	&	0.99	&	40.0	&	12597	&	0.117	&	1/1	&		\\	\hline
248861930	&	12.93	$\pm$	0.26	&	23.353	&	1.00	&	28.0	&	36903	&	0.521	&	1/1	&	DBA	\\	\hline
247992671	&	15.4	$\pm$	0.4	&	8.742	&	1.00	&	10.8	&	8000	&	0.23	&	1/1	&		\\	\hline
241256067	&	15.778	$\pm$	0.028	&	5.352	&	1.00	&	6.0	&	4730	&	0.495	&	1/2	&	DC	\\	\hline
124894242	&	0.32555	$\pm$	0.00017	&	4.075	&	1.00	&	6.9	&	15919	&	0.435	&	1/1	&		\\	\hline
154942231	&	4.4236	$\pm$	0.0022	&	7.839	&	1.00	&	6.4	&	12000	&	0.15	&	3/5	&		\\	\hline
649932200	&	0.9712	$\pm$	0.0005	&	31.107	&	1.00	&	28.4	&	26461	&	1.036	&	3/3	&		\\	\hline
52584399	&	3.801	$\pm$	0.008	&	112.873	&	1.00	&	66.8	&	19505	&	0.249	&	3/3	&		\\	\hline
1883129777	&	1.9446	$\pm$	0.0004	&	143.936	&	1.00	&	71.9	&	16480	&	0.229	&	3/3	&		\\	\hline
1881564044	&	6.137	$\pm$	0.004	&	6.042	&	1.00	&	7.4	&	34000	&	0.2	&	2/3	&	DAH \\	\hline
1101460688	&	23.24	$\pm$	0.03	&	10.905	&	1.00	&	6.9	&	88000	&	0.656	&	2/3	&		\\	\hline
22014729	&	7.6477	$\pm$	0.0032	&	135.408	&	1.00	&	55.8	&	36580	&	0.44	&	1/2	&		\\	\hline
31798074	&	21.828 $\pm$ 0.017	&42.369	&	1.00	&	5.3	&	12011	&	0.398	&	1/10	&		\\	\hline
262921847	&	13.892	$\pm$	0.004	&	52.290	&	1.00	&	20.7	&	7000	&	0.23	&	12/12	&		\\	\hline
311886800	&	10.2157	$\pm$	0.0025	&	32.576	&	1.00	&	19.4	&	100000	&	0.6	&	6/6	&		\\	\hline
178994042	&	0.2138	$\pm$	7e-05	&	4.400	&	1.00	&	5.6	&	19770	&	0.57	&	1/1	&		\\	\hline
220478684	&	16.611 $\pm$ 0.010	&	25.143	&	1.00	&	12.6	&	44589	&	0.483	&	19/20	&		\\	\hline
381976323	&	8.722	$\pm$	0.01	&	133.547	&	1.00	&	45.2	&	4000	&	0.2	&	11/11	&		\\	\hline
235072034	&	84.0	$\pm$	4.0	&	20.540	&	1.00	&	10.5	&	3500	&	0.23	&	2/3	&		\\	\hline
219244444	&	7.2886	$\pm$	0.0027	&	245.106	&	1.00	&	12.5	&	6440	&	0.25	&	3/5	&	DA7+M+planet	\\	\hline
245834374	&	33.6	$\pm$	1.8	&	7.275	&	1.00	&	6.8	&	34210	&	0.69	&	1/1	&	DA+M	\\	\hline
85146644	&	5.317	$\pm$	0.0011	&	68.977	&	1.00	&	51.4	&	60260	&	0.41	&	3/3	&		\\	\hline
107310223	&	11.28	$\pm$	0.23	&	21.505	&	1.00	&	30.9	&	48116	&	0.366	&	1/1	&	DA1+dM	\\	\hline
1101900758	&	2.65191	$\pm$	0.0003	&	31.335	&	1.00	&	21.4	&	23941	&	0.398	&	5/5	&		\\	\hline
160119816	&	21.453	$\pm$	0.017	&	112.794	&	1.00	&	43.3	&	90000	&	0.56	&	9/9	&		\\	\hline
300284179	&	12.65	$\pm$	0.29	&	12.426	&	1.00	&	13.8	&	38490	&	0.54	&	1/1	&		\\	\hline
126910998	&	19.4	$\pm$	0.6	&	36.753	&	1.00	&	54.9	&	23075	&	0.819	&	1/1	&	DA+M	\\	\hline
218971976	&	6.7081	$\pm$	0.001	&	263.431	&	1.00	&	65.1	&	39260	&	0.38	&	2/2	&	DA+M	\\	\hline
321979116	&	2.69506	$\pm$	0.00016	&	92.402	&	1.00	&	36.5	&	8650	&	0.96	&	2/2	&	DAP5.8	\\	\hline
183533908	&	2.55267	$\pm$	0.00024	&	11.122	&	1.00	&	13.6	&	7638	&	0.74	&	1/1	&		\\	\hline
630539472	&	1.592	$\pm$	0.004	&	4.231	&	1.00	&	7.4	&	10027	&	0.821	&	1/1	&		\\	\hline
201255204	&	6.34	$\pm$	0.03	&	59.895	&	1.00	&	52.0	&	4000	&	0.2	&	2/2	&	DBH	\\	\hline
439917321	&	12.98	$\pm$	0.28	&	61.269	&	1.00	&	48.6	&	19193	&	0.57	&	1/1	&	DA+M	\\	\hline
55096188	&	2.75096	$\pm$	0.00021	&	4.379	&	1.00	&	6.5	&	10000	&	0.6	&	2/3	&		\\	\hline
53190694	&	9.923	$\pm$	0.004	&	12.529	&	1.00	&	9.3	&	30502	&	0.41	&	3/4	&	DA3+dM	\\	\hline
952264940	&	1.78894	$\pm$	0.00018	&	7.713	&	1.00	&	8.6	&	25692	&	1.151	&	1/2	&	DBH	\\	\hline
219769847	&	8.72	$\pm$	0.13	&	13.789	&	1.00	&	20.7	&	21620	&	0.72	&	1/1	&	DA3+dM	\\	\hline
292114384	&	0.29255	$\pm$	0.00014	&	4.110	&	1.00	&	5.1	&	18943	&	0.57	&	1/1	&		\\	\hline
66714896	&	12.579	$\pm$	0.009	&	10.848	&	1.00	&	8.8	&	63073	&	0.69	&	1/3	&		\\	\hline
349233389	&	52.0	$\pm$	5.0	&	31.595	&	1.00	&	25.7	&	23690	&	0.58	&	1/1	&	DA2+M	\\	\hline
335501964	&	6.7654	$\pm$	0.0025	&	5.380	&	1.00	&	8.1	&	90000	&	0.53	&	1/2	&		\\	\hline
165916724	&	0.092325	$\pm$	5e-07	&	4.156	&	1.00	&	5.2	&	9557	&	0.883	&	1/3	&	DAH	\\	\hline
274039489	&	7.32	$\pm$	0.09	&	67.842	&	1.00	&	58.5	&	42370	&	0.66	&	1/1	&	DA1.2+pair	\\	\hline
328029653	&	14.157	$\pm$	0.004	&	19.335	&	0.99	&	12.5	&	6340	&	0.53	&	5/5	&		\\	\hline
311920919	&	0.1614889	$\pm$	1e-06	&	4.611	&	1.00	&	6.2	&	14116	&	0.68	&	1/4	&		\\	\hline
155871645	&	0.1385659	$\pm$	8e-07	&	6.552	&	1.00	&	5.5	&	34050	&	0.77	&	3/5	&		\\	\hline
344927121	&	2.53418	$\pm$	0.00018	&	113.339	&	1.00	&	67.3	&	26088	&	0.381	&	4/4	&		\\	\hline
148719841	&	8.549	$\pm$	0.004	&	17.920	&	1.00	&	18.7	&	6169	&	0.72	&	2/2	&		\\	\hline
415714190	&	0.384926	$\pm$	4e-06	&	9.333	&	1.00	&	8.4	&	7690	&	0.636	&	3/3	&		\\	\hline
277713491	&	8.9986	$\pm$	0.0032	&	11.795	&	1.00	&	9.2	&	7518	&	0.578	&	3/4	&		\\	\hline
471013515	&	6.05	$\pm$	0.07	&	27.262	&	1.00	&	34.8	&	55000	&	0.6	&	1/1	&		\\	\hline
5393020	&	1.5786	$\pm$	0.00014	&	54.426	&	1.00	&	49.4	&	8000	&	0.35	&	1/2	&		\\	\hline
610754879	&	0.91146	$\pm$	9e-05	&	11.175	&	1.00	&	10.9	&	47617	&	1.341	&	2/3	&		\\	\hline
365207316	&	2.539	$\pm$	0.005	&	6.316	&	1.00	&	7.1	&	10520	&	0.226	&	2/2	&		\\	\hline
425079955	&	17.076	$\pm$	0.016	&	20.821	&	1.00	&	16.2	&	14252	&	0.285	&	2/2	&		\\	\hline
1271382181	&	7.426	$\pm$	0.005	&	40.083	&	1.00	&	21.8	&	24015	&	0.25	&	8/8	&		\\	\hline
115970839	&	2.6994	$\pm$	0.0004	&	41.932	&	1.00	&	30.0	&	8825	&	0.219	&	5/5	&	DA+M:	\\	\hline
159819681	&	10.728	$\pm$	0.006	&	29.967	&	1.00	&	22.8	&	27876	&	0.573	&	5/5	&		\\	\hline
158971558	&	6.327	$\pm$	0.004	&	7.300	&	1.00	&	7.3	&	14319	&	0.323	&	2/2	&		\\	\hline
1550807820	&	0.566897	$\pm$	1.6e-05	&	5.636	&	1.00	&	8.6	&	44523	&	1.293	&	5/5	&		\\	\hline
7622101	&	8.5708	$\pm$	0.0028	&	11.742	&	1.00	&	7.6	&	14159	&	0.2	&	7/7	&		\\	\hline
349993999	&	4.381	$\pm$	0.0005	&	7.642	&	1.00	&	7.0	&	24751	&	0.756	&	1/1	&		\\	\hline
1688534307	&	0.0902002	$\pm$	4e-07	&	4.253	&	1.00	&	5.3	&	66522	&	0.637	&	1/2	&		\\	\hline
62918265	&	4.6612	$\pm$	0.0008	&	12.845	&	0.99	&	13.1	&	8133	&	0.724	&	6/6	&		\\	\hline
1922433415	&	2.03387	$\pm$	0.00015	&	129.416	&	1.00	&	60.4	&	21155	&	0.214	&	2/2	&		\\	\hline
100469915	&	12.936	$\pm$	0.006	&	18.310	&	1.00	&	21.6	&	12644	&	0.152	&	2/2	&		\\	\hline
2026050241	&	8.0844	$\pm$	0.0024	&	6.393	&	1.00	&	5.7	&	48831	&	0.489	&	1/2	&		\\	\hline
60112260	&	7.37	$\pm$	0.1	&	14.436	&	1.00	&	20.0	&	7682	&	0.756	&	1/1	&		\\	\hline
269937578	&	7.2969	$\pm$	0.0019	&	79.671	&	1.00	&	45.4	&	32755	&	0.274	&	4/5	&		\\	\hline
2052430471	&	0.125404	$\pm$	2.9e-05	&	4.277	&	1.00	&	5.4	&	28678	&	0.368	&	1/1	&		\\	\hline
124573902	&	7.9918	$\pm$	0.0035	&	21.546	&	1.00	&	21.9	&	12173	&	0.371	&	2/2	&	wd+pair	\\	\hline
270432557	&	9.311	$\pm$	0.01	&	6.316	&	1.00	&	5.4	&	10000	&	0.23	&	2/3	&		\\	\hline
43529091	&	9.44	$\pm$	0.14	&	11.479	&	1.00	&	14.0	&	11006	&	0.17	&	1/1	&		\\	\hline
278861557	&	3.1624	$\pm$	0.0004	&	26.520	&	1.00	&	7.8	&	10460	&	0.23	&	19/20	&	wd+pair	\\	\hline
62846288	&	0.500906	$\pm$	1.4e-05	&	23.831	&	1.00	&	17.9	&	16268	&	0.935	&	3/4	&		\\	\hline
56813164	&	8.05	$\pm$	0.11	&	25.988	&	1.00	&	34.7	&	17304	&	0.39	&	1/1	&	DA+M	\\	\hline
953086708	&	0.088138	$\pm$	1.3e-05	&	5.293	&	1.00	&	7.4	&	7735	&	0.564	&	1/1	&	DAEH	\\	\hline
95475692	&	1.94319	$\pm$	0.00021	&	45.117	&	1.00	&	36.2	&	13737	&	0.428	&	2/2	&	DA+M	\\	\hline
68015843	&	3.015	$\pm$	0.0018	&	21.609	&	1.00	&	21.7	&	12392	&	0.4	&	2/2	&	DA+M4.5	\\	\hline
219485299	&	1.9770 $\pm$ 0.0020	&	4.513	&	1.00	&	18.9	&	10307	&	0.63	&	3/3	&	DA+M	\\	\hline
1507548590	&	1.9272	$\pm$	0.0004	&	50.124	&	1.00	&	49.2	&	25089	&	0.214	&	2/2	&		\\	\hline
1101327387	&	1.30545	$\pm$	9e-05	&	21.739	&	1.00	&	19.5	&	15659	&	1.13	&	2/2	&	DAH	\\	\hline
416538823	&	16.021	$\pm$	0.011	&	335.044	&	1.00	&	26.1	&	18840	&	0.8	&	4/4	&	DA+M	\\	\hline
7983187	&	14.32	$\pm$	0.16	&	77.693	&	1.00	&	45.2	&	17210	&	0.59	&	2/2	&	DA3+dM	\\	\hline
251857373	&	2.31675	$\pm$	0.00012	&	175.807	&	1.00	&	110.6	&	40698	&	0.56	&	2/2	&		\\	\hline
408015814	&	29.52	$\pm$	0.05	&	13.089	&	1.00	&	11.8	&	76010	&	0.061	&	2/2	&		\\	\hline
200725303	&	9.48	$\pm$	0.07	&	35.024	&	1.00	&	33.5	&	90000	&	0.75	&	2/2	&		\\	\hline
159894126	&	0.46415	$\pm$	0.00017	&	4.410	&	1.00	&	5.9	&	8627	&	0.613	&	1/2	&		\\	\hline
651923454	&	42.4	$\pm$	2.9	&	16.656	&	1.00	&	11.9	&	21644	&	0.746	&	1/1	&	wd+pair	\\	\hline
651337335	&	30.5	$\pm$	1.5	&	6.331	&	1.00	&	9.7	&	21796	&	0.217	&	1/1	&		\\	\hline
650999219	&	24.6	$\pm$	0.5	&	24.641	&	1.00	&	26.3	&	70000	&	0.49	&	2/2	&		\\	\hline
93031595	&	9.94	$\pm$	0.16	&	17.306	&	1.00	&	25.7	&	8751	&	0.911	&	1/1	&	DZ	\\	\hline
685012668	&	2.8735	$\pm$	0.0018	&	107.688	&	1.00	&	44.9	&	20041	&	0.238	&	5/5	&		\\	\hline
672383331	&	17.8	$\pm$	0.5	&	7.107	&	1.00	&	10.9	&	21076	&	0.279	&	1/1	&		\\	\hline
257689590	&	13.35	$\pm$	0.13	&	5.132	&	0.99	&	5.2	&	8629	&	0.183	&	2/2	&		\\	\hline
455072135	&	11.36	$\pm$	0.21	&	90.850	&	1.00	&	49.0	&	12965	&	0.116	&	1/1	&	EB	\\	\hline
941411508	&	10.547	$\pm$	0.006	&	8.630	&	1.00	&	9.1	&	40000	&	0.556	&	2/2	&		\\	\hline
942842236	&	8.233	$\pm$	0.004	&	10.210	&	1.00	&	10.1	&	77164	&	0.75	&	2/2	&		\\	\hline
332828037	&	2.07518	$\pm$	0.00024	&	15.973	&	1.00	&	17.4	&	13936	&	0.382	&	2/2	&		\\	\hline
939953390	&	29.85	$\pm$	0.05	&	9.949	&	1.00	&	6.7	&	63208	&	0.667	&	3/3	&		\\	\hline
952000432	&	21.158	$\pm$	0.025	&	23.145	&	1.00	&	23.4	&	81500	&	0.58	&	2/2	&		\\	\hline
232524621	&	3.4918	$\pm$	0.0005	&	14.341	&	1.00	&	8.5	&	100000	&	0.65	&	2/9	&	hotDA+pair	\\	\hline
1003414425	&	26.46	$\pm$	0.04	&	25.074	&	1.00	&	16.5	&	18089	&	0.173	&	4/4	&		\\	\hline
233177285	&	15.194	$\pm$	0.01	&	4.790	&	1.00	&	6.2	&	6500	&	0.2	&	2/6	&		\\	\hline
232979174	&	8.8045	$\pm$	0.0033	&	8.642	&	1.00	&	5.6	&	12776	&	0.473	&	2/9	&		\\	\hline
1001602651	&	4.9125	$\pm$	0.0018	&	10.704	&	1.00	&	7.2	&	9786	&	0.826	&	5/6	&		\\	\hline
219009693	&	1.5828	$\pm$	0.0004	&	9.973	&	1.00	&	7.3	&	14000	&	0.23	&	4/4	&		\\	\hline
364160098	&	0.19744	$\pm$	6e-05	&	4.307	&	1.00	&	5.7	&	20000	&	0.2	&	1/1	&		\\	\hline
471013541	&	0.549	$\pm$	0.0005	&	4.071	&	1.00	&	5.6	&	88642	&	0.51	&	1/1	&		\\	\hline
951282189	&	2.9618	$\pm$	0.0005	&	95.469	&	0.99	&	110.8	&	19099	&	0.174	&	2/2	&	DAB+pair	\\	\hline
951635495	&	23.45	$\pm$	0.03	&	7.002	&	1.00	&	6.6	&	61045	&	0.5	&	2/2	&		\\	\hline
1924161836	&	25.132	$\pm$	0.024	&	14.278	&	1.00	&	12.4	&	71080	&	0.482	&	2/2	&		\\	\hline
410414842	&	14.9	$\pm$	0.4	&	6.417	&	1.00	&	9.7	&	70000	&	0.5	&	1/1	&		\\	\hline
2052284133	&	16.652	$\pm$	0.03	&	12.980	&	1.00	&	14.3	&	34299	&	0.51	&	2/2	&		\\	\hline
251080865	&	32.01	$\pm$	0.06	&	179.265	&	1.00	&	38.0	&	7170	&	0.744	&	2/2	&	DCP7	\\	\hline
164681986	&	3.9437	$\pm$	0.0009	&	33.642	&	1.00	&	31.3	&	8471	&	1.006	&	2/2	&		\\	\hline
392797216	&	0.6808	$\pm$	0.0004	&	6.037	&	1.00	&	6.0	&	6814	&	0.749	&	2/2	&	DAH	\\	\hline
157201137	&	0.776229	$\pm$	3.3e-05	&	10.198	&	1.00	&	10.9	&	11254	&	0.486	&	2/2	&	CV	\\	\hline
219868627	&	5.9071	$\pm$	0.0017	&	43.787	&	1.00	&	18.3	&	18000	&	0.18	&	12/12	&	DAZ+pair	\\	\hline
155872634	&	3.9145	$\pm$	0.0007	&	26.280	&	1.00	&	16.7	&	24589	&	0.697	&	6/6	&		\\	\hline
357389336	&	42.46	$\pm$	0.08	&	7.007	&	1.00	&	6.1	&	21785	&	0.49	&	2/5	&	DA+K	\\	\hline
313894558	&	19.05	$\pm$	0.04	&	4.861	&	0.99	&	5.2	&	18610	&	0.61	&	1/2	&		\\	\hline
441569276	&	2.47353	$\pm$	0.00023	&	90.759	&	1.00	&	55.4	&	19690	&	0.43	&	6/6	&	DA+M	\\	\hline
457099062	&	1.36615	$\pm$	0.0002	&	7.658	&	1.00	&	8.2	&	6655	&	0.555	&	2/2	&		\\	\hline
198510602	&	8.87	$\pm$	0.013	&	110.918	&	1.00	&	46.3	&	9800	&	0.23	&	4/4	&		\\	\hline
236865474	&	17.789	$\pm$	0.011	&	55.795	&	1.00	&	21.0	&	26520	&	0.52	&	14/14	&	DA1.9+pair	\\	\hline
279484490	&	0.086195	$\pm$	1.2e-05	&	4.519	&	1.00	&	5.8	&	16989	&	0.387	&	1/1	&		\\	\hline
248353420	&	11.32	$\pm$	0.26	&	105.385	&	1.00	&	69.0	&	60419	&	0.53	&	1/1	&	DA+K	\\	\hline
54003343	&	22.864	$\pm$	0.029	&	16.612	&	1.00	&	18.5	&	52869	&	0.611	&	1/2	&	DA1+dM	\\	\hline
71513592	&	23.0	$\pm$	0.9	&	11.500	&	1.00	&	15.8	&	39910	&	0.438	&	1/1	&		\\	\hline
737660462	&	9.703	$\pm$	0.013	&	30.331	&	1.00	&	23.7	&	30415	&	0.502	&	5/5	&		\\	\hline
60040774	&	9.72	$\pm$	0.15	&	9.104	&	1.00	&	14.7	&	8876	&	0.277	&	1/1	&	wd+pair	\\	\hline
712210226	&	0.115822	$\pm$	2.2e-05	&	4.178	&	1.00	&	5.5	&	13707	&	0.53	&	1/1	&	wd+pair	\\	\hline
705867935	&	0.28689	$\pm$	0.00013	&	6.605	&	1.00	&	13.7	&	16407	&	1.001	&	1/1	&		\\	\hline
765410943	&	3.7702	$\pm$	0.0005	&	169.012	&	1.00	&	55.6	&	22601	&	0.238	&	19/19	&		\\	\hline
96391732	&	3.662	$\pm$	0.011	&	98.438	&	1.00	&	117.0	&	21272	&	0.183	&	2/2	&		\\	\hline
291303146	&	14.36	$\pm$	0.33	&	43.492	&	1.00	&	40.9	&	100000	&	0.58	&	1/1	&		\\	\hline
770913638	&	0.453289	$\pm$	1.1e-05	&	19.139	&	1.00	&	21.2	&	8474	&	0.794	&	2/2	&		\\	\hline
743328948	&	7.9	$\pm$	0.1	&	28.663	&	1.00	&	45.6	&	20000	&	0.2	&	1/1	&		\\	\hline
403672198	&	3.137	$\pm$	0.016	&	25.892	&	1.00	&	38.6	&	7750	&	0.23	&	1/1	&		\\	\hline
437213912	&	37.1	$\pm$	2.3	&	29.331	&	1.00	&	34.2	&	9000	&	0.2	&	1/1	&		\\	\hline
836046041	&	7.9019	$\pm$	0.0034	&	12.734	&	1.00	&	15.5	&	26187	&	0.517	&	2/2	&		\\	\hline
817911025	&	0.597463	$\pm$	1.9e-05	&	17.981	&	1.00	&	16.0	&	84000	&	0.63	&	3/4	&		\\	\hline
832516183	&	30.55	$\pm$	0.05	&	9.190	&	1.00	&	9.2	&	28583	&	0.403	&	1/1	&		\\	\hline
804549027	&	2.948	$\pm$	0.007	&	117.121	&	1.00	&	79.0	&	10000	&	0.2	&	2/2	&		\\	\hline
191532802	&	3.553	$\pm$	0.022	&	5.513	&	1.00	&	11.6	&	8630	&	0.808	&	1/1	&		\\	\hline
262708252	&	5.4411	$\pm$	0.0015	&	28.747	&	1.00	&	16.1	&	10266	&	0.152	&	7/7	&		\\	\hline
841325183	&	2.504	$\pm$	0.0005	&	10.370	&	1.00	&	9.2	&	17205	&	0.28	&	3/4	&		\\	\hline
873285117	&	20.79	$\pm$	0.024	&	9.782	&	1.00	&	9.9	&	37453	&	0.349	&	1/2	&		\\	\hline
875311151	&	11.053	$\pm$	0.007	&	9.792	&	1.00	&	8.7	&	50475	&	0.491	&	1/2	&		\\	\hline
309025727	&	8.22	$\pm$	0.06	&	19.948	&	1.00	&	26.6	&	9000	&	0.23	&	1/2	&		\\	\hline
870310469	&	23.443	$\pm$	0.03	&	9.701	&	1.00	&	9.9	&	37491	&	0.502	&	2/2	&		\\	\hline
841399917	&	36.32	$\pm$	0.15	&	64.304	&	1.00	&	39.3	&	17000	&	0.2	&	4/4	&		\\	\hline
33357141	&	13.47	$\pm$	0.31	&	8.387	&	1.00	&	11.5	&	24952	&	0.526	&	1/1	&		\\	\hline
345036441	&	1.64579	$\pm$	8e-05	&	13.629	&	1.00	&	10.6	&	10355	&	0.92	&	4/5	&	DAH	\\	\hline
471013677	&	2.79285	$\pm$	0.00031	&	59.858	&	1.00	&	53.6	&	36470	&	0.6	&	1/4	&		\\	\hline
902670676	&	22.988	$\pm$	0.029	&	10.934	&	1.00	&	10.8	&	78000	&	0.66	&	2/2	&		\\	\hline
82347011	&	6.2108	$\pm$	0.0021	&	130.009	&	1.00	&	94.0	&	30071	&	0.4	&	2/2	&	DA+M3	\\	\hline
950218375	&	10.3	$\pm$	0.19	&	7.101	&	1.00	&	10.0	&	42180	&	0.52	&	1/1	&		\\	\hline
232972481	&	81.13	$\pm$	0.28	&	89.769	&	1.00	&	39.0	&	99575	&	0.87	&	7/7	&	DA+M	\\	\hline
156073827	&	0.189861	$\pm$	2e-06	&	10.486	&	1.00	&	10.0	&	29300	&	0.415	&	2/3	&		\\	\hline
115613388	&	0.703919	$\pm$	2.6e-05	&	6.117	&	1.00	&	5.3	&	28169	&	1.272	&	2/5	&	DAH	\\	\hline
630123582	&	4.109 $\pm$ 0.027	&	3.846	&	1.00	&	12.5	&	60000	&	0.456	&	1/1	&		\\	\hline
630314437	&	0.17219	$\pm$	5e-05	&	4.425	&	1.00	&	6.9	&	21128	&	0.526	&	1/1	&	wd+pair	\\	\hline
399570361	&	10.89	$\pm$	0.22	&	25.327	&	0.99	&	27.6	&	79000	&	0.656	&	1/1	&	wd+pair	\\	\hline
183799565	&	0.689793	$\pm$	2.6e-05	&	5.825	&	1.00	&	5.8	&	180000	&	0.54	&	1/4	&		\\	\hline
142982488	&	0.40188	$\pm$	0.00026	&	4.112	&	1.00	&	5.6	&	34180	&	0.626	&	1/1	&		\\	\hline
298411553	&	2.90014	$\pm$	0.00031	&	12.166	&	1.00	&	8.2	&	39229	&	0.756	&	4/8	&		\\	\hline
80858168	&	2.84122	$\pm$	0.0003	&	38.679	&	1.00	&	34.0	&	12891	&	0.256	&	3/3	&		\\	\hline
652008669	&	48.0	$\pm$	4.0	&	6.901	&	1.00	&	6.0	&	31092	&	0.432	&	1/1	&		\\	\hline
412252434	&	2.40744 $\pm$ 0.00022	&	7.758	&	1.00	&	12.8	&	10206	&	0.597	&	4/4	&		\\	\hline
760481896	&	6.641	$\pm$	0.005	&	64.546	&	1.00	&	32.8	&	27691	&	0.404	&	4/4	&		\\	\hline
453006983	&	9.093	$\pm$	0.005	&	19.789	&	1.00	&	21.5	&	27425	&	0.52	&	2/2	&	DA+M	\\	\hline
804378446	&	29.44	$\pm$	0.05	&	12.366	&	1.00	&	14.6	&	48986	&	0.685	&	2/2	&		\\	\hline
310478036	&	2.36875 $\pm$ 0.00029	&	11.001	&	1.00	&	59.3	&	7000	&	0.23	&	7/7	&		\\	\hline
349408306	&	3.6493	$\pm$	0.0005	&	84.357	&	1.00	&	25.5	&	8000	&	0.15	&	21/21	&		\\	\hline
462519068	&	4.987 $\pm$ 0.008	&	4.490	&	1.00	&	19.7	&	7000	&	0.15	&	3/3	&		\\	\hline
119736060	&	3.954	$\pm$	0.026	&	49.116	&	1.00	&	56.6	&	22600	&	0.42	&	1/1	&	DA+M	\\	\hline
46199750	&	1.891	$\pm$	0.008	&	7.440	&	1.00	&	13.9	&	20000	&	0.94	&	1/1	&	DAP	\\	\hline
147918835	&	2.722	$\pm$	0.012	&	156.826	&	1.00	&	63.5	&	24661	&	0.21	&	1/1	&		\\	\hline
142808656	&	7.9038	$\pm$	0.0034	&	11.463	&	1.00	&	10.0	&	18646	&	0.031	&	4/4	&		\\	\hline
258086867	&	30.3	$\pm$	1.5	&	7.607	&	1.00	&	5.6	&	7573	&	0.753	&	1/1	&		\\	\hline
1000946479	&	2.69774 $\pm$ 0.00032	&	4.572	&	1.00	&	5.8	&	7139	&	0.6	&	1/4	&	DC	\\	\hline
274928438	&	2.464	$\pm$	0.01	&	5.552	&	1.00	&	6.6	&	6959	&	0.23	&	1/1	&	DA+M	\\	\hline
1100582330	&	2.736	$\pm$	0.017	&	18.131	&	1.00	&	29.2	&	26828	&	0.496	&	1/1	&		\\	\hline
284900652	&	7.9923	$\pm$	0.0021	&	359.311	&	1.00	&	71.7	&	34733	&	0.36	&	29/29	&	DA+dMe	\\	\hline
383673264	&	72.37	$\pm$	0.17	&	15.671	&	1.00	&	6.8	&	11435	&	0.661	&	11/25	&		\\	\hline
25771075	&	7.8738	$\pm$	0.0023	&	14.170	&	1.00	&	14.5	&	15849	&	0.263	&	3/3	&		\\	\hline
47499442	&	7.83	$\pm$	0.1	&	18.473	&	1.00	&	22.7	&	7184	&	0.562	&	1/1	&		\\	\hline
33692053	&	15.1	$\pm$	0.4	&	14.656	&	1.00	&	20.4	&	24809	&	0.239	&	1/1	&		\\	\hline
34855316	&	4.461	$\pm$	0.032	&	30.907	&	1.00	&	46.5	&	8021	&	0.745	&	1/1	&		\\	\hline
355154138	&	2.37013	$\pm$	0.00029	&	18.259	&	1.00	&	11.4	&	14569	&	0.403	&	7/7	&		\\	\hline
429233827	&	7.17	$\pm$	0.0028	&	113.682	&	1.00	&	74.9	&	10618	&	0.174	&	2/2	&		\\	\hline
32159409	&	42.8	$\pm$	3.2	&	9.008	&	1.00	&	11.7	&	10000	&	0.2	&	1/1	&		\\	\hline
1550940367	&	21.89	$\pm$	0.19	&	21.620	&	1.00	&	15.7	&	50000	&	0.425	&	4/4	&		\\	\hline
91193988	&	3.6933	$\pm$	0.0005	&	18.393	&	1.00	&	19.9	&	13597	&	0.249	&	2/2	&		\\	\hline
192991819	&	1.81005	$\pm$	0.00018	&	34.905	&	1.00	&	35.9	&	7000	&	0.23	&	3/3	&	wd+pair	\\	\hline
73764818	&	0.135393	$\pm$	3.2e-05	&	4.101	&	1.00	&	5.0	&	35994	&	0.464	&	1/1	&		\\	\hline
178539314	&	4.3923	$\pm$	0.0011	&	5.228	&	1.00	&	7.8	&	8201	&	0.422	&	2/2	&		\\	\hline
938779482	&	0.4613	$\pm$	0.0004	&	8.293	&	0.96	&	14.2	&	10000	&	0.2	&	1/1	&	wd+pair	\\	\hline
251903434	&	34.6	$\pm$	1.9	&	109.202	&	0.98	&	28.1	&	7752	&	0.702	&	1/1	&		\\	\hline
159394587	&	14.753	$\pm$	0.021	&	7.390	&	0.85	&	5.2	&	16890	&	0.6	&	1/4	&		\\	\hline
74342209	&	1.557	$\pm$	0.004	&	146.008	&	0.87	&	44.0	&	9463	&	0.478	&	1/1	&		\\	\hline
1201364356	&	276.0	$\pm$	6.0	&	483.909	&	0.86	&	20.9	&	12816	&	0.61	&	11/15	&	wd+pair-disk?	\\	\hline
398365709	&	8.0828	$\pm$	0.0018	&	72.725	&	0.92	&	19.0	&	22570	&	0.54	&	5/5	&	DA+M	\\	\hline
245830829	&	13.56	$\pm$	0.15	&	300.596	&	0.96	&	18.7	&	17620	&	0.63	&	2/2	&	DA3+M	\\	\hline
329670050	&	2.703	$\pm$	0.011	&	99.642	&	0.98	&	34.9	&	9000	&	0.23	&	1/1	&		\\	\hline
243349100	&	23.447	$\pm$	0.028	&	14.703	&	0.98	&	8.6	&	7250	&	0.23	&	5/6	&		\\	\hline
1920685932	&	25.12	$\pm$	0.06	&	40.975	&	0.99	&	12.5	&	11981	&	0.388	&	1/2	&	wd+pair	\\	\hline
269663822	&	4.5085	$\pm$	0.0008	&	226.862	&	0.98	&	75.2	&	7000	&	0.2	&	20/20	&		\\	\hline
403292348	&	3.3522	$\pm$	0.0012	&	280.702	&	0.98	&	101.4	&	8000	&	0.2	&	2/2	&		\\	\hline
15131739	&	6.143	$\pm$	0.004	&	17.562	&	0.99	&	16.5	&	31194	&	0.403	&	2/2	&	DA+L3	\\	\hline
142616553	&	181.0	$\pm$	1.7	&	31.440	&	0.84	&	19.0	&	14635	&	0.48	&	4/6	&	wd+pair	\\	\hline
901406326	&	0.113414	$\pm$	2.2e-05	&	5.396	&	0.87	&	9.5	&	114711	&	0.701	&	1/1	&		\\	\hline
115013365	&	1.20234 $\pm$ 0.00005	&	6.454	&	1.00	&	9.0	&	12826	&	0.23	&	1/2	&		\\	\hline
269071459	&	21.51	$\pm$	0.025	&	17.219	&	0.98	&	10.5	&	9375	&	0.852	&	5/5	&		\\	\hline
137607346	&	23.927	$\pm$	0.025	&	190.904	&	0.98	&	58.7	&	9000	&	0.2	&	6/6	&		\\	\hline
458692929	&	27.4	$\pm$	1.2	&	5.070	&	0.84	&	8.2	&	35105	&	0.36	&	1/1	&	DA+pair	\\	\hline
311948770	&	16.932	$\pm$	0.01	&	219.451	&	0.80	&	89.7	&	49650	&	0.66	&	5/5	&	DAO+M	\\	\hline
903362877	&	65.4	$\pm$	1.3	&	14.507	&	0.98	&	5.2	&	18344	&	0.488	&	1/3	&	wd+pair	\\	\hline
158747696	&	2.041	$\pm$	0.007	&	23.570	&	0.98	&	27.8	&	8000	&	0.205	&	1/1	&	wd+pair	\\	\hline
8158475	&	56.85	$\pm$	0.18	&	8.819	&	0.99	&	6.9	&	45333	&	0.527	&	1/2	&		\\	\hline
1990690286	&	15.322	$\pm$	0.009	&	17.081	&	1.00	&	16.1	&	100000	&	0.48	&	2/2	&		\\	\hline
814547613	&	3.3579	$\pm$	0.0006	&	27.264	&	0.96	&	19.8	&	14115	&	0.567	&	2/4	&		\\	\hline
88509155	&	128.0	$\pm$	27.0	&	28.056	&	1.00	&	29.4	&	56610	&	0.53	&	1/1	&	DA+M	\\	\hline
1951569775	&	92.9	$\pm$	0.9	&	237.305	&	1.00	&	36.3	&	7000	&	0.2	&	2/2	&		\\	\hline
115075496	&	53.0	$\pm$	4.0	&	9.833	&	1.00	&	9.9	&	44715	&	0.469	&	1/1	&	DA+M:	\\	\hline
118690168	&	34.85	$\pm$	0.07	&	7.436	&	1.00	&	6.2	&	11023	&	0.587	&	2/3	&		\\	\hline
138653638	&	213.5	$\pm$	2.6	&	40.343	&	0.78	&	17.0	&	12436	&	0.478	&	3/4	&	wd+pair	\\	\hline
721486224	&	150.0	$\pm$	50.0	&	150.795	&	0.84	&	17.1	&	11692	&	0.725	&	3/3	& wd+pair	\\	\hline
1101853877	&	121.0	$\pm$	11.0	&	102.164	&	1.00	&	7.0	&	18791	&	0.673	&	1/2	&	DA+pair	\\	\hline
453444667	&	82.1	$\pm$	1.5	&	111.345	&	1.00	&	59.4	&	105000	&	0.75	&	2/2	&	DA+BP	\\	\hline
246847488	&	173.0	$\pm$	5.0	&	51.455	&	1.00	&	21.5	&	5720	&	0.1	&	1/1	&		\\	\hline
262548040	&	2* (0.10614391	$\pm$	2.5e-7)	&	6.409	&	1.00	&	6.2	&	6230	&	0.41	&	3/5	&	DA7H	\\	\hline
229098638	&	181.6	$\pm$	0.7	&	815.491	&	1.00	&	19.2	&	18800	&	0.56	&	4/4	&	wd+pair	\\	\hline
404156391	&	28.133	$\pm$	0.030	&	221.896	&	0.97	&	8.8	&	17770	&	0.542	&	2/2	&	wd+pair	\\	\hline
204440456	&	6.20	$\pm$	0.07	&	18.679	&	1.00	&	21.3	&	6908	&	0.6018	&	1/1	&		\\	\hline
732235000	&	23.667	$\pm$	0.020	&	19.728	&	1.00	&	8.5	&	19212	&	0.523	&	11/16	&	wd+pair	\\	\hline
50385872	&	5.4166	$\pm$	0.0007	&	5.535	&	1.00	&	5.6	&	13141	&	0.710	&	2/8	&		\\	\hline
61965938	&	6.8168	$\pm$	0.0013	&	51.961	&	1.00	&	5.1	&	19505	&	0.633	&	1/3	&		\\	\hline
1174761001	&	5.7114	$\pm$	0.0018	&	23.495	&	1.00	&	24.2	&	18311	&	0.511	&	2/2	&	wd+pair	\\	\hline
23936802	&	5.8054	$\pm$	0.0018	&	17.076	&	1.00	&	27.0	&	17146	&	0.269	&	4/4	&	\\	\hline	
253936074	&	45.16	$\pm$	0.07	&	25.890	&	1.00	&	5.5	& 29347 	& 0.676	&	1/2	&	wd+pair \\	\hline		
2055541164	&	2.53003	$\pm$	0.00023	&	73.753	&	1.00	&	48.6	&	23575	&	0.261	&	4/4	&	\\	\hline	
408253347	&	7.9829	$\pm$	0.0034	&	49.765	&	1.00	&	36.4	&	24466	&	0.47	&	2/3	&	DA+dMe	\\	\hline
611305959	&	21.84	$\pm$	0.05	&	21.142	&	1.00	&	16.4	&	90000	&	0.6	&	3/3	&	\\	\hline	
611853501	&	0.385874	$\pm$	0.000006	&	13.347	&	1.00	&	14.9	&	21697	&	1.055	&	2/2	&	\\	\hline	
2024143938	&	1.1367	$\pm$	0.0023	&	17.506	&	1.00	&	29.1	&	20000	&	0.323	&	1/1	&	DAZ+L3	\\	\hline
380174982	&	4.4379	$\pm$	0.0005	&	15.711	&	0.98	&	15.3	&	8607	&	0.83	&	3/3	&	\\	\hline	
219099282	&	77.17 $\pm$ 0.20	&	22.276	&	1.00	&	6.3	&	45230	&	0.63	&	15/25	&	\\	\hline	
610721330	&	0.671865	$\pm$	0.000017	&	5.779	&	1.00	&	5.0	&	24332	&	1.218	&	1/4	&	\\	\hline	
161820334	&	6.9232	$\pm$	0.002	&	32.422	&	1.00	&	15.3	&	18462	&	0.183	&	6/7	&	\\	\hline	
441702898	&	8.3222	$\pm$	0.0029	&	13.601	&	0.99	&	8.6	&	7501	&	0.445	&	6/6	&	\\	\hline	
144197100	&	0.245758	$\pm$	0.000008	&	16.037	&	1.00	&	16.5	&	8595	&	0.87	&	2/2	&	\\	\hline	 
841424790	&	3.311	$\pm$	0.0004	&	7.040	&	1.00	&	6.0	&	73027	&	0.751	&	3/7	&	\\	\hline
308292831 & 4.61 $\pm$ 0.04 & 6.218 & 1.00 & 12.0 & 9230 & 0.191 & 1/1 & ELMV+pair \\ \hline  
256351952 & 28.868 $\pm$ 0.029 & 47.669 & 1.00 & 20.3 & 10403 & 0.172 & 5/5 &   \\ \hline
900086743 & 2.26444 $\pm$ 0.00029 & 55.902 & 1.00 & 58.0 & 23230 & 0.39 & 2/2 &   \\ \hline
370238674 & 3.3448 $\pm$ 0.0004 & 34.483 & 1.00 & 22.4 & 9453 & 0.788 & 6/6 &   \\ \hline
337219837 & 6.8813 $\pm$ 0.0018 & 175.005 & 1.00 & 70.4 & 25203 & 0.441 & 6/6 & DA+M-B \\ \hline
802015620 & 12.218 $\pm$ 0.006 & 73.835 & 1.00 & 58.6 & 12982 & 0.263 & 3/3 & wd+pair \\ \hline
742167700 & 18.778 $\pm$ 0.013 & 17.158 & 1.00 & 14.9 & 80000 & 0.6 & 3/3 & DO \\ \hline
358247592 & 5.7849 $\pm$ 0.0012 & 10.705 & 1.00 & 8.7 & 33601 & 0.572 & 4/4 & wd+pair \\ \hline
468604087 & 3.5365 $\pm$ 0.0005 & 18.791 & 1.00 & 17.6 & 147000 & 0.706 & 7/7 & wd+pair \\ \hline
232065830 & 4.5928 $\pm$ 0.0005 & 87.323 & 1.00 & 49.8 & 19211 & 0.511 & 7/7 &   \\ \hline
437042755 & 5.7082 $\pm$ 0.0017 & 31.903 & 1.00 & 23.8 & 18981 & 1.06 & 4/4 & DBH \\ \hline
321741942 & 24.16 $\pm$ 0.022 & 24.560 & 1.00 & 27.9 & 84000 & 0.82 & 2/2 &   \\ \hline
274239484 & 1.92716 $\pm$ 0.00013 & 41.484 & 1.00 & 23.0 & 7743 & 0.764 & 9/9 & DAH \\ \hline
362023614 & 12.627 $\pm$ 0.006 & 21.236 & 1.00 & 16.1 & 53220 & 0.56 & 3/3 &   \\ \hline
471015333 & 18.817 $\pm$ 0.019 & 13.115 & 1.00 & 14.7 & 70000 & 0.49 & 2/2 &   \\ \hline
620324891 & 10.7947 $\pm$ 0.0032 & 30.343 & 1.00 & 19.5 & 70000 & 0.634 & 5/5 &   \\ \hline
32307067 & 1.90674 $\pm$ 0.0002 & 36.376 & 1.00 & 33.5 & 16500 & 0.39 & 2/2 &   \\ \hline
140937916 & 3.3278 $\pm$ 0.0005 & 27.793 & 1.00 & 20.8 & 15084 & 0.112 & 4/4 &   \\ \hline
407569944 & 19.185 $\pm$ 0.014 & 28.815 & 1.00 & 23.6 & 9955 & 0.552 & 3/3 &   \\ \hline
155577443 & 39.5 $\pm$ 2.5 & 12.277 & 1.00 & 17.1 & 4000 & 0.6 & 1/1 &   \\ \hline
840458841 & 6.79 $\pm$ 0.07 & 5.110 & 1.00 & 6.2 & 58333 & 1.105 & 1/1 &   \\ \hline
397979837 & 2.67088 $\pm$ 0.00031 & 68.242 & 1.00 & 46.4 & 31810 & 0.37 & 4/4 &   \\ \hline
150251751 & 53.85 $\pm$ 0.16 & 13.089 & 1.00 & 13.0 & 7145 & 0.722 & 2/2 &   \\ \hline
171630076 & 3.399 $\pm$ 0.018 & 52.974 & 1.00 & 56.1 & 15000 & 0.93 & 1/1 & DAP-B \\ \hline
377056497 & 1.5254 $\pm$ 0.00012 & 8.006 & 1.00 & 6.0 & 10060 & 1.06 & 3/3 & DAH-B \\ \hline
357686299 & 31.57 $\pm$ 0.033 & 18.783 & 1.00 & 8.5 & 49240 & 0.63 & 10/10 & DA \\ \hline
321159503 & 0.465797 $\pm$ 0.000008 & 8.501 & 1.00 & 7.3 & 7750 & 0.8 & 3/3 & DAH \\ \hline
741451769 & 0.187876 $\pm$ 0.000004 & 10.426 & 1.00 & 13.0 & 30190 & 1.33 & 2/2 & DC \\ \hline
313144249 & 0.723464 $\pm$ 0.000019 & 47.048 & 1.00 & 27.9 & 10671 & 0.845 & 8/8 &   \\ \hline
611402948 & 2.24076 $\pm$ 0.00019 & 76.558 & 1.00 & 63.7 & 30000 & 0.2 & 2/2 & wd+pair \\ \hline
91329200 & 1.69439 $\pm$ 0.00008 & 10.872 & 1.00 & 6.9 & 9146 & 0.88 & 5/5 &   \\ \hline
339451801 & 106.4 $\pm$ 0.5 & 50.761 & 1.00 & 14.8 & 8742 & 0.236 & 2/2 & wd+pair \\ \hline
117502568 & 65.27 $\pm$ 0.16 & 6.337 & 1.00 & 6.3 & 97080 & 0.53 & 1/1 &   \\ \hline
20656977 & 6.2179 $\pm$ 0.0011 & 8.339 & 1.00 & 10.1 & 14065 & 0.35 & 2/2 &   \\ \hline
62823020 & 1.3488 $\pm$ 0.001 & 18.161 & 1.00 & 15.8 & 8380 & 0.164 & 3/3 &   \\ \hline
66419115 & 7.4439 $\pm$ 0.0012 & 54.261 & 1.00 & 44.6 & 7711 & 0.307 & 3/3 &   \\ \hline
620251540 & 82 $\pm$ 10 & 136.482 & 1.00 & 20.4 & 15425 & 0.53 & 1/1 & wd+pair \\ \hline
270679102 & 3.5858 $\pm$ 0.0007 & 71.393 & 1.00 & 42.7 & 17505 & 0.37 & 3/3 & DA+M \\ \hline
103222871 & 2.28912 $\pm$ 0.00019 & 38.290 & 0.98 & 21.7 & 8597 & 0.76 & 6/6 &   \\ \hline
91898005 & 1.0509 $\pm$ 0.00006 & 18.931 & 1.00 & 14.9 & 7765 & 0.881 & 3/3 &   \\ \hline
417179225 & 19.194 $\pm$ 0.02 & 131.832 & 0.84 & 42.4 & 22260 & 0.62 & 2/2 & wd+pair \\ \hline
23992223 & 16.627 $\pm$ 0.015 & 11.080 & 1.00 & 10.5 & 15160 & 0.47 & 2/2 & wd+pair \\ \hline
141117787 & 6.7091 $\pm$ 0.0017 & 14.991 & 1.00 & 16.3 & 6802 & 0.914 & 2/2 & DAH \\ \hline
311629054 & 19.2 $\pm$ 0.6 & 12.252 & 1.00 & 14.5 & 19780 & 0.62 & 1/1 &   \\ \hline
406782194 & 1.83857 $\pm$ 0.00019 & 4.955 & 1.00 & 5.2 & 27800 & 0.87 & 1/1 &   \\ \hline
54636377 & 25.633 $\pm$ 0.018 & 19.880 & 1.00 & 15.7 & 19440 & 0.58 & 2/2 &   \\ \hline
53709985 & 8.362 $\pm$ 0.0026 & 31.571 & 1.00 & 23.8 & 10466 & 0.126 & 3/3 &   \\ \hline
201892746 & 4.0085 $\pm$ 0.0006 & 13.392 & 1.00 & 12.4 & 6476 & 0.746 & 2/2 & DAH \\ \hline
22704782 & 21.503 $\pm$ 0.025 & 6.093 & 1.00 & 5.5 & 21144 & 0.5 & 2/2 &   \\ \hline
299437785 & 2.05802 $\pm$ 0.00016 & 10.585 & 0.98 & 12.4 & 8553 & 0.815 & 1/1 &   \\ \hline
840874557 & 0.20919 $\pm$ 0.00007 & 4.218 & 0.91 & 5.4 & 24966 & 0.599 & 1/1 & wd+pair \\ \hline
91903603 & 0.1237488 $\pm$ 0.0000009 & 4.484 & 1.00 & 5.4 & 6880 & 0.868 & 1/2 &   \\ \hline

\end{longtable}

\bibliography{references}{}
\bibliographystyle{aasjournal}

\end{document}